\pdfoutput=1
\documentclass[fleqn,10pt]{article}
\usepackage[utf8]{inputenc}
\usepackage{ifthen}
\usepackage{amsmath}
\usepackage{amssymb}
\usepackage{amsfonts}
\usepackage{amscd}
\usepackage{cite}
\usepackage{accents}
\usepackage{mathrsfs}
\usepackage[footnotesize,hang]{subfigure}
\usepackage{cancel} % \cancel{stuff} provides slashed stuff
\usepackage{longtable}
\usepackage[footnotesize,bf]{caption}
\usepackage{url}
\usepackage{bbold}
\usepackage{bm}% bold greek letters using \bm{}
\usepackage[punctsep]{collref}
\usepackage{multirow}
\usepackage[T1]{fontenc}

\usepackage{graphicx}
\usepackage{hyperref}

\hypersetup{
pdfauthor={Martin Holthausen, Manfred Lindner and Michael A. Schmidt},
pdftitle={Radiative Symmetry Breaking of the Minimal Left-Right Symmetric Model}
}

\newcommand{\tr}{\ensuremath{\mathrm{tr}}}

\newcommand{\I}{\ensuremath{\mathrm{i}}}
\newcommand{\D}{\ensuremath{\mathrm{d}}}

\newcommand{\GeV}{\ensuremath{\,\mathrm{GeV}}}
\newcommand{\TeV}{\ensuremath{\,\mathrm{TeV}}}

\newcommand{\braket}[1]{\ensuremath{\left<#1\right>}}
\newcommand{\hc}{\ensuremath{\text{h.c.}}}
\newcommand{\Order}[1]{\ensuremath{\mathcal{O}\left(#1\right)}}

\newcommand{\SU}[1]{\ensuremath{\mathrm{SU}(#1)}}
\newcommand{\Spin}[1]{\ensuremath{\mathrm{Spin}(#1)}}
\newcommand{\U}[1]{\ensuremath{\mathrm{U}(#1)}}
\newcommand{\SO}[1]{\ensuremath{\mathrm{SO}(#1)}}

\newcommand{\Rep}[1]{\ensuremath{\underline{\mathbf{#1}}}}

\newcommand{\ko}[1]{\ensuremath{\mathcal{O}\left(#1\right)}}

\newcommand{\Eqref}[1]{Eq. (\ref{#1})}
\newcommand{\Eqqref}[2]{Eqs. (\ref{#1}, \ref{#2})}

\newcommand{\Figref}[1]{Fig. \ref{#1}}

\newcommand{\Tabref}[1]{Tab. \ref{#1}}

\newcommand{\Secref}[1]{Sec. \ref{#1}}
\newcommand{\Seccref}[2]{Secs. \ref{#1} and \ref{#2}}
\newcommand{\Appref}[1]{App. \ref{#1}}

\newcommand{\spin}[2]{\ensuremath{\left(\Rep{#1},\Rep{#2}\right)}}

\newcommand{\VEV}[1]{\langle #1 \rangle}

\newcommand{\Lp}{\ensuremath{\mathbb{L}}}
\newcommand{\Qu}{\ensuremath{\mathbb{Q}}}

\newcommand{\Bd}{\ensuremath{\mathbb{\Phi}}}
\newcommand{\Dt}{\ensuremath{\mathbb{\Psi}}}

\newcommand{\oLp}{\ensuremath{L_L}}
\newcommand{\oQu}{\ensuremath{Q_L}}
\newcommand{\oLpD}{\ensuremath{L_R}}
\newcommand{\oQuD}{\ensuremath{Q_R}}

\newcommand{\oBd}{\ensuremath{\Phi}}
\newcommand{\oBdD}{\ensuremath{\tilde\Phi^{\dagger}}}
\newcommand{\oDt}{\ensuremath{\chi_L}}
\newcommand{\oDtD}{\ensuremath{\chi_R}}

\allowdisplaybreaks[1]

\newcommand{\MSbar}{\ensuremath{\overline{\mathrm{MS}}\;}}

\newcommand{\AccSymm}{\ensuremath{\mathcal{A}_\Dt}}
\newcommand{\andone}{\hspace{1 cm} \mathrm{and} \hspace{1 cm}}

\begin{document}
\begin{titlepage}
\ \vspace*{-15mm}
\begin{flushright}
IPPP/09/88\\
DCPT/09/176
\end{flushright}
\vspace*{5mm}

\begin{center}
{\Huge\sffamily\bfseries
Radiative Symmetry Breaking of the Minimal Left-Right Symmetric Model
}
\\[10mm]
{\large
Martin Holthausen\footnote{martin.holthausen@mpi-hd.mpg.de}$^{(a)}$,
Manfred Lindner\footnote{lindner@mpi-hd.mpg.de}$^{(a)}$ and
Michael A. Schmidt\footnote{m.a.schmidt@dur.ac.uk}$^{(b)}$}
\\[5mm]
{
\small\textit{$^{(a)}$ Max--Planck--Institut f\"{u}r Kernphysik, Saupfercheckweg 1, 69117 Heidelberg, Germany
}}
\\[3mm]
{\small\textit{$^{(b)}$
Institute for Particle Physics Phenomenology, University of Durham, Durham, DH1 3LE, UK
}}
\\[5mm]
\end{center}
\vspace*{1.0cm}

\begin{abstract}
\noindent  Under the assumption of classical conformal invariance, we study the Coleman-Weinberg symmetry breaking mechanism in the minimal left-right symmetric model. This model is attractive as it provides a natural framework for small neutrino masses and the restoration of parity as a good symmetry of nature. We find that, in a large fraction of the parameter space, the parity symmetry is maximally broken by quantum corrections in the Coleman-Weinberg potential, which are a consequence of the conformal anomaly. As the left-right symmetry breaking scale is connected to the Planck scale through the logarithmic running of the dimensionless couplings of the scalar potential, a large separation of the two scales can be dynamically generated. The symmetry breaking dynamics of the model was studied using a renormalization group analysis. Electroweak symmetry breaking is triggered by the breakdown of left-right symmetry, and the left-right breaking scale is therefore expected in the few TeV range. The phenomenological implications of the symmetry breaking mechanism are discussed.
\end{abstract}

\end{titlepage}

\newpage
\setcounter{footnote}{0}

\section{Introduction}
The Standard Model (SM) of particle physics has been explaining a wide variety of
experimental observations and even its quantum nature has been established by
e.g.~the correct prediction of the top mass by its imprint on electroweak
precision observables. The most unsatisfying part of the SM, however, is given
by its description of electroweak symmetry breaking via the Higgs mechanism. As the mass term of the Higgs field is quadratically sensitive to the physics
at the cutoff scale $\Lambda$, its smallness vis-a-vis e.g.~the Planck scale or any
other high energy scale poses a severe \emph{naturalness} problem.
Furthermore, the origin of the large hierarchy between the Planck scale $M_P$ and
  the electroweak scale $\mu$ remains unexplained.
This has motivated many extensions of the SM such as models based on
supersymmetry, which explains the stability and the electroweak scale is
  linked to the supersymmetry scale, as well as extra dimensions, where
  the fundamental (4+N)-dimensional Planck scale is lowered.

The Higgs mass term as the only super-renormalizable operator in the SM lies at the heart of the hierarchy problem. Since it is also the only dimensionful parameter in the SM, its absence leads to classical conformal invariance of the SM. It has been argued by Bardeen~\cite{Bardeen} that once classical conformal invariance is imposed on the SM, the quadratic divergences appear as unphysical manifestations of the chosen regularization procedure and disappear by the use the anomalous Ward identity of conformal invariance, which ensures a minimal breaking of conformal symmetry by the anomaly. Of course, this can not be considered as a solution of the hierarchy problem, as this argument cannot be invoked for embeddings of the SM in an underlying quantum field theory.

Recently, however, this idea was reconsidered by Meissner and Nicolai~\cite{Meissner:2006zh,Meissner:2007xv,Meissner:2009gs}, who argued that classical conformal invariance of the particle physics action might be a consequence of a finite quantum gravitational embedding at the Planck scale. The logarithmic contributions to the effective action, which are proportional to the beta functions and thus lie at the heart of the anomaly, are reinterpreted as the leading quantum gravitational effects in the particle physics action. Electroweak symmetry breaking would then be triggered by Planck scale effects. While the generation of the hierarchy of the two scales can be explained by the logarithms which communicate the conformal breaking down from the Planck scale physics, the stabilization of the electroweak scale is due to assumption that the Wilsonian argument does not apply to the Planck scale.  If one should call this a solution to the hierarchy problem, we leave up to the reader.

Electroweak precision measurements and results from flavor physics seem to indicate that the physics of electroweak symmetry breaking might be less rich than initially hoped for and we are thus motivated to study more minimalistic proposals. Note, however, that there are a number of criteria that a classically conformally invariant theory has to fulfill, if one wants to embed it at the Planck scale in this way: there should be no Landau poles up to the Planck scale; the Higgs potential should remain stable up to the Planck scale; and there can not be any intermediate scale.
Before we discuss a realization of this idea in the context of low-scale left-right symmetry, let us first briefly review the Coleman-Weinberg symmetry breaking mechanism. In their classic study of massless scalar quantum electrodynamics, they showed that the conformal invariance is broken by quantum corrections to the effective potential. If one chooses the renormalization group scale such that the tree-level potential vanishes, the Coleman-Weinberg potential lifts the flat direction and the scalar mass is one-loop suppressed with respect to the gauge boson mass (and calculable!).

In the CW calculation, perturbative reliability of the calculation requires the one-loop
gauge contribution to the effective potential to be smaller than the tree level
term. The tree level potential thus has to be sufficiently flat which can be
achieved by a suitable renormalization point. The logarithmic
renormalization group (RG) running can then naturally explain a hierarchy between the initial scale and the symmetry
breaking scale. In the scale-invariant SM, however, due to the large top mass
$m_t>m_Z$ the effective potential is rendered unstable~\cite{Lindner:1985uk,Lindner:1988ww} and the SM thus has to be
extended. As new scalar and vector degrees of freedom give positive
contributions to the effective potential it is not surprising that e.g for
shadow~\cite{Sher:1988mj,Hempfling:1996ht,Casas:2000mn,Nishino:2004kb,Meissner:2006zh,Espinosa:2007qk,Chang:2007ki,Foot:2007as,Foot:2007ay,Hambye:2007vf,Meissner:2008gj}
and other~\cite{Iso:2009ss,Iso:2009nw} extensions of the Higgs sector this
problem can be circumvented and a successful phenomenology can be achieved.\\

In this work instead of adding singlets to the SM we discuss conformal
invariance in the context of the minimal left-right symmetric model based on the
gauge group $\SU{2}_L\times\SU{2}_R\times\U{1}_{B-L}\times\SU{3}_C$
~\cite{Mohapatra:1974hk,Senjanovic:1975rk} that has been long known as an
attractive extension of the SM as it explains parity violation by spontaneous
symmetry breaking and has a natural place for neutrino masses. For simplicity,
we restrict ourselves to the simplest version of the minimal LR symmetric model
(including a $\mathbb{Z}_4$ symmetry~\cite{Adulpravitchai:2009re}) which
contains, in addition to the bidoublet, a pair of doublets that are used to
break LR symmetry. In this context, neutrinos are Dirac particles. Majorana
neutrinos are obtained either via the seesaw
mechanism~\cite{Minkowski:1977sc,Yanagida:1980,Glashow:1979vf,Gell-Mann:1980vs,Mohapatra:1980ia}
if there are triplets~\cite{Mohapatra:1979ia} instead of
doublets or by an inverse seesaw
mechanism~\cite{Mohapatra:1986aw,Mohapatra:1986bd} if there is an additional
scalar singlet and at least two fermionic singlets.\\
As the one-loop effective (Coleman-Weinberg) potential of a gauge theory with an
extended Higgs sector --such as the minimal LR symmetric model we are
considering here-- contains contributions from several mass scales, there are
multiple logarithms with different arguments which complicate the
minimization. However, the one loop contribution within the perturbative regime
becomes only important in parameter space regions of a small tree-level potential.
Therefore, we use the method of Gildener and Weinberg
(GW)~\cite{Gildener:1976ih} in the discussion of the Higgs potential.\\
The outline of the paper is as follows. In \Secref{sec:SBofParity}, we outline the model, review the GW method and discuss
how parity is spontaneously broken in the simplified Higgs potential containing
doublets only. In order to discuss the RG flow, the beta functions have been
calculated and the relevant ones are presented. In \Secref{sec:RealisticModel},
we include the bidoublet in the discussion, at first, as a perturbation and
subsequently, we discuss the Higgs potential of bidoublet and
doublets. Furthermore, we discuss fermion masses and the bounds from flavor
changing neutral currents (FCNCs). Finally, we summarize and draw our
conclusions in \Secref{sec:summary}. The beta functions of the minimal LR
symmetric model as well as other details are summarized in the appendix.

\section{Spontaneous Breaking of Parity}
\label{sec:SBofParity}
In this chapter, we first review the GW method to investigate the
parity  breaking in scale invariant scalar potentials. Then we apply it to the
minimal LR symmetric potential and discuss how parity is broken. Finally we
demonstrate how the GW conditions are achieved by renormalization group (RG) evolution.

\subsection{Gildener Weinberg Method in the Minimal LR Symmetric Potential}
\label{sec:SBonlyDoublets}
As parity is a symmetry of the left-right symmetric model, we use the
isomorphism \mbox{$\SU{2}\times\SU{2}\cong\Spin{4}$} to express all fields in
terms of representations of $\Spin{4}$, which is described by the Clifford
algebra of $\SO{4}$\footnote{Mathematical details of the
  $\Spin{4}$ group are summarized in \Appref{app:spin4}.}.
In this simplified notation, the bidoublet $\Bd$ is represented by a complex vector
representation and the additional left- and right-handed Higgs doublets $\oDt$, $\oDtD$, as well
as the SM fermions, form $\Spin{4}$ Dirac spinors. The
  particle content of the minimal left-right symmetric model is shown in
\Tabref{tab:partcontspinor}.
\begin{table}
\begin{center}
\begin{tabular}{c|c|c|c}
  particle & parity $\mathcal{P}$ & $\mathbb{Z}_4$  & $\Spin{1,3}\times(\SU{2}_L\times\SU{2}_R)\times\left(\SU{3}_C\times\U{1}_{B-L}\right)$ \\\hline
  $\Lp_{1,2,3}=\left(\begin{array}{c}\oLp\\-\I\oLpD\end{array}\right)$ &
  $P\mathbb{P}\Lp(t,-x)$  & $L_R\rightarrow \I L_R$ & $\left[\spin{\frac12}{0}\spin{2}{1}+\spin{0}{\frac12}\spin{1}{2}\right]\left(\Rep{1},-1\right)$ \\
  $\Qu_{1,2,3}=\left(\begin{array}{c}\oQu\\-\I\oQuD\end{array}\right)$ & $P\mathbb{P}\Qu(t,-x)$ & $Q_R\rightarrow -\I Q_R$  &
   $\left[\spin{\frac12}{0}\spin{2}{1}+\spin{0}{\frac12}\spin{1}{2}\right]\left(\Rep{3},\frac13\right)$
   \\
\hline
  $\Bd=\left(\begin{array}{cc} 0 &
      \oBd \\ -\oBdD & 0\end{array}\right)$ & $\mathbb{P}\Bd^\dagger\mathbb{P}(t,-x)$ &  $\Bd\rightarrow \I\Bd$  &
 $\spin{0}{0}\spin{2}{2}\left(\Rep{1},0\right)$ \\
  $\Dt=\left(\begin{array}{c}\oDt \\-\I \oDtD\end{array}\right)$ & $\mathbb{P}\Dt(t,-x)$ & $\oDtD\rightarrow -\I\oDtD$  &
$\spin{0}{0}\left[\spin{2}{1}+\spin{1}{2}\right]\left(\Rep{1},-1\right)$ \\
\end{tabular}
\caption{Particle content in spinor notation. $\oBd$ is the bidoublet in the
  notation of~\cite{Deshpande:1990ip} and $\tilde\oBd=\sigma_2\oBd^*\sigma_2$ is
  the charge conjugate. $P=\gamma^0$ and $\mathbb{P}=\Gamma^4$ denote the
  Lorentz group and $\Spin{4}$ parity matrices, respectively. The decomposition in physical fields and
  the complete Lagrangian are presented in \Appref{app:Lagrangian}.\label{tab:partcontspinor}}
\end{center}
\end{table}
In the $\Spin{4}$ notation, the most general scale- and gauge-invariant scalar
potential\footnote{The complete Lagrangian is shown in \Appref{app:Lagrangian}.} respecting the $\mathrm{Z}_4$ symmetry is given by:
\begin{equation}
\begin{split}
\mathcal{V}(\Bd,\Dt)&
=\frac{\kappa_1}{2}\left(\overline{\Dt}\Dt\right)^2+\frac{\kappa_2}{2}
  \left(\overline{\Dt}\Gamma\Dt\right)^2
+\lambda_1\left(\tr\Bd^{\dagger}\Bd\right)^2
+\lambda_2\left(\tr\Bd\Bd+\tr\Bd^{\dagger}\Bd^{\dagger}\right)^2
+\lambda_3\left(\tr\Bd\Bd-\tr\Bd^{\dagger}\Bd^{\dagger}\right)^2\\
&+\beta_1\, \overline{\Dt}\Dt\tr\Bd^\dagger\Bd
+f_1\,\overline{\Dt}\Gamma [\Bd^\dagger,\Bd]\Dt\;,
\label{eq:VHiggssolo}
\end{split}
\end{equation}
where $\Gamma=\Gamma^1\Gamma^2\Gamma^3\Gamma^4$ denotes the chirality operator of $\Spin{4}$ and the $\Gamma^A$ form a representation of the Clifford algebra of $\Spin{4}$.
As all couplings are real, the Higgs potential is CP conserving. The separate $\Spin{4}$ transformations of $\Dt$ and $\Bd$ are broken to
the diagonal subgroup, unless the coupling $f_1$ vanishes. Note, the operator of
$\beta_1$ can be rewritten as  \mbox{$\overline{\Dt}\Dt \tr\Bd^\dagger \Bd=2\overline{\Dt}\{\Bd^\dagger,\Bd\}\Dt$}.
 Due to the conformal symmetry, the dimension three term
  $\overline{\Dt}\Bd\Dt$ is not allowed and thus there is an accidental symmetry
  \mbox{$\AccSymm: \Dt\rightarrow \mathrm{e}^{\I\beta\Gamma}\Dt$} whose
  implications are discussed in more detail in \Secref{sec:fermionmasses}.

As we assume the theory to be weakly coupled, quantum corrections can be taken into
account by a loop expansion of the effective potential.  We will consider
  the effective potential up to one loop. In order to discuss
symmetry breaking, we have to minimize the potential. However, even the
minimization of the one-loop effective potential cannot be done analytically in
the case of multiple scalar fields. Instead of resorting to a numerical study,
we will use the analytical approximate method of Gildener and Weinberg~\cite{Gildener:1976ih} which makes essential use of the renormalization group.  \\
GW have noted that for a generic scale invariant potential of the form
\begin{align}
  V_0=\frac{1}{24}f_{ijkl}\Phi_i\Phi_j\Phi_k\Phi_l
\end{align}
the renormalization group can be used to enforce a single condition on the scalar couplings of the theory:
\begin{align}
\underset{N_i N_i=1}{\min} \left( f_{ijkl}(\mu_{GW})N_i N_j N_k N_l \right)\Big|_{N_i=n_i}=0.
\label{eq:GWcondition}
\end{align}
This condition entails that at the scale $\mu_{GW}$, the scalar potential has a
tree level \emph{flat direction} $\Phi_i=n_i \phi$. Barring the possibility of
accidental additional flat directions, radiative corrections dominate in this
direction in field space while they can be neglected in all other directions (see~\cite{Gildener:1976ih}). In the \MSbar scheme, the one-loop effective potential in the flat direction $\Phi=n\, \phi$ can be easily calculated~\cite{Coleman:1973jx} to be
 \begin{align}
 \delta V(n \phi)=A\phi^4+B \phi^4 \ln \frac{\phi^2}{\mu_{GW}^2}
\end{align}
with
\begin{subequations}
\begin{align}
A&=\frac{1}{64\pi^2{\VEV{\phi}}^4}\sum_{i}n_i M_i^4(n\VEV{\phi})\left( \ln \frac{M_i^2(n\VEV{\phi})}{{\VEV{\phi}}^2} -c_i\right)  \\
B&=\frac{1}{64\pi^2{\VEV{\phi}}^4}    \sum_{i}n_i M_i^4(n\VEV{\phi}) \;,
\end{align}%
\label{eq:AandB}%
\end{subequations}%
where $n_i$ denotes the degrees of freedom, $M_i$ is the mass and $c_i=\frac32$
for scalars and fermions and $c_i=\frac56$ for gauge bosons.
The stationary condition $\left.\frac{\partial \delta V_{\rm{1-loop}}}{\partial
  \phi}\right|_{\phi=\VEV{\phi}}=0$ results in
\begin{equation}\label{eq:mincond}
\ln \frac{\VEV{\phi}^2}{\mu_{GW}^2}=-\frac12 -\frac AB
\end{equation}
and the mass of the excitation in the flat direction -- so called \emph{scalon}
$s$, which is the \emph{pseudo-Nambu Goldstone boson} (pNGB) of broken scale invariance -- is given by
\begin{align}
M_S^2= n_i n_j \left.\frac{\partial^2\delta
  V(n\phi)}{\partial\phi_i\partial\phi_j}\right|_{n\braket{\phi}}=\left.\frac{\D^2}{\D\phi^2}V(n\phi)\right|_{\braket{\phi}}=
8 B\braket{\phi}^2=\frac{1}{8\pi^2\braket{\phi}^2}\left(\tr
  M_S^4+3\tr M_V^4-4\tr M_D^4\right)\label{eq:scalonmass}.
\end{align}
From  \Eqref{eq:GWcondition}, we see that the application of the
GW method in the context of the minimal left-right symmetric
potential requires the minimization of this very complicated potential on a unit
sphere in field space.
Parameterizing the scalar fields as
\begin{align}
\Dt=\frac{1}{\sqrt{2}}\left(
\begin{array}{c}
N_1 e^{\I \theta} \\
N_5 e^{\I \vartheta_5}\\
N_2 e^{\I \vartheta_2} \\
N_6 e^{\I \vartheta_6}
\end{array}
\right)\phi
\andone
 \oBd =\frac{1}{2}\left(
\begin{array}{cc}
 N_3 e^{\I \vartheta_3} & N_7 e^{\I \vartheta_7} \\
 N_8 e^{\I \vartheta_8} & N_4  e^{\I \alpha}\\
\end{array}
\right)\phi\;,
\end{align}
the GW conditions read
\begin{align}
\sum_i n_i^2=1 , \hspace{1cm}
\mathcal{V}\Big|_{N_i=n_i}=0\andone
\frac{\partial}{\partial N_i}\mathcal{V}\Big|_{N_i=n_i}=0
\label{eq:GWchap6}
\end{align}
where $\{n_i\}$ parameterizes a flat direction of the left-right symmetric
potential. The first condition normalizes the VEVs such that the sum of squares
lies on a unit sphere. The second condition ensures that the potential vanishes along the flat direction. Finally, the third set of relations is the
condition of the flat direction to be actually an extremum. Hence, we are
minimizing the Higgs potential with two constraints.

We restrict ourselves to the case where the electromagnetic gauge
group is left unbroken as required by phenomenology. Hence the VEV
configuration of a certain flat direction is given by
\begin{align}
\langle\Dt\rangle=\left(
\begin{array}{c}
v_L e^{\I \theta} \\
 0 \\
v_R\\
 0
\end{array}
\right)=
\frac{1}{\sqrt{2}}\left(
\begin{array}{c}
n_1 e^{\I \theta} \\
 0 \\
n_2\\
 0
\end{array}
\right)\VEV{\phi}
\quad\text{and}\quad
 \langle\oBd\rangle =\frac{1}{\sqrt{2}}\left(
\begin{array}{cccc}
 \kappa & 0 \\
 0 & \kappa^\prime  e^{\I \alpha} \\
\end{array}
\right)=
\frac{1}{2}\left(
\begin{array}{cccc}
 n_3 & 0 \\
 0 & n_4  e^{\I \alpha} \\
\end{array}
\right)\VEV{\phi}.
\end{align}
We have used the gauge freedom to set the phases
$\vartheta_2=\vartheta_3=0$.
The relation to the VEVs defined is given by
\begin{align}
v_L&=n_1 \frac{\VEV{\phi}}{\sqrt{2}}    &v_R&=n_2 \frac{\VEV{\phi}}{\sqrt{2}}   &\kappa&=n_3 \frac{\VEV{\phi}}{\sqrt{2}}   &  \kappa^\prime&=n_4 \frac{\VEV{\phi}}{\sqrt{2}} .
\end{align}
The explicit form of the Gildener-Weinberg conditions is given in \Eqref{eq:GWconditionsfull}. In the
remainder of this section, we discuss the limit of vanishing vacuum expectation
values for the bidoublet and discuss the RG evolution.
In \Secref{sec:RealisticModel} we discuss the resulting multi-Higgs doublet
model at low-energies and investigate the full Higgs potential.

\subsection{LR Symmetry Breaking in the Limit of Vanishing Bidoublet VEVs}
\label{sec:flatdirectionsofdoublet}
Since phenomenology requires the right-handed VEV $v_R$ to be much larger than the
electroweak scale, it is prudent to neglect the bidoublet in a first step and
consider the GW mechanism in the Higgs potential containing only the doublets:
\begin{equation*}
\mathcal{V}_\Dt=\frac{\kappa_1}{2}
\left(\overline{\Dt}\Dt\right)^2+\frac{\kappa_2}{2} \left(\overline{\Dt}\Gamma\Dt\right)^2.
\end{equation*}
Then, in a second step, we will treat the bidoublet potential in the LR-broken
phase. In unitary gauge, the potential at the minimum reads
\begin{equation}
 \mathcal{V}_\Dt\Big|_{N_i=n_i}=\frac{1}{8}\phi^4\left(\kappa_1+\kappa_2(-1+2
   n_1^2)^2 \right)\; ,
\end{equation}
where we have used the normalization condition of the VEVs to eliminate $n_2$.
 Stability of this potential requires it to be bounded from below. The term
 multiplying $\phi^4$ therefore has to be positive semi-definite in the entire range of
 $n_1$. Inserting the minimum and boundary values $n_1^2=1/2$ and $n_1^2=1$, respectively, we find the stability conditions $\kappa_1\geq0$ and $\kappa_+=\kappa_1+\kappa_2\geq0$.\\
The GW conditions after the insertion of the normalization condition\footnote{We
  drop the factor $\phi^n$ in the GW conditions.}
\begin{subequations}
\begin{align}
 0=&\left.\frac{\partial}{\partial N_1}\mathcal{V}_\Dt\right|_{N_i=n_i}=\frac12 n_1 \left(\kappa_1+\kappa_2\left(-1+2 n_1^2\right)\right)\\
 0=&\left.\frac{\partial}{\partial N_2}\mathcal{V}_\Dt\right|_{N_i=n_i}=\frac12 \sqrt{1-n_1^2} \left(\kappa_1-\kappa_2\left(-1+2 n_1^2\right)\right)\\
0=&\mathcal{V}_\Dt\Big|_{N_i=n_i}=\frac18\left(\kappa_1+\kappa_2(-1+2 n_1^2)^2 \right)
\end{align}
\end{subequations}
allow for three flat directions
\begin{enumerate}
\item $n_1=1,\;n_2=0,\;\kappa_+=0$,
\item $n_1=0,\; n_2=1,\; \kappa_+=0$ and
\item $n_1=n_2=\frac{1}{\sqrt{2}}\; ,\kappa_1=0$\; .
\end{enumerate}
The first two correspond to parity breaking and are phenomenologically equivalent,
since in the unbroken phase there is no difference between left- and right-handed
fields and we can define the direction which acquires a VEV as the right-handed
one. Therefore, we do not discuss the second flat direction in the following.
If $\kappa_2$ is negative, there are only maximally left-right symmetry breaking
flat directions.

\subsection{Renormalization Group Evolution}
\label{sec:RGparity}
\begin{figure}\centering
\subfigure[RG flow without gauge  boson
contributions. \label{fig:kappa1}]{\includegraphics[width=7.5 cm]{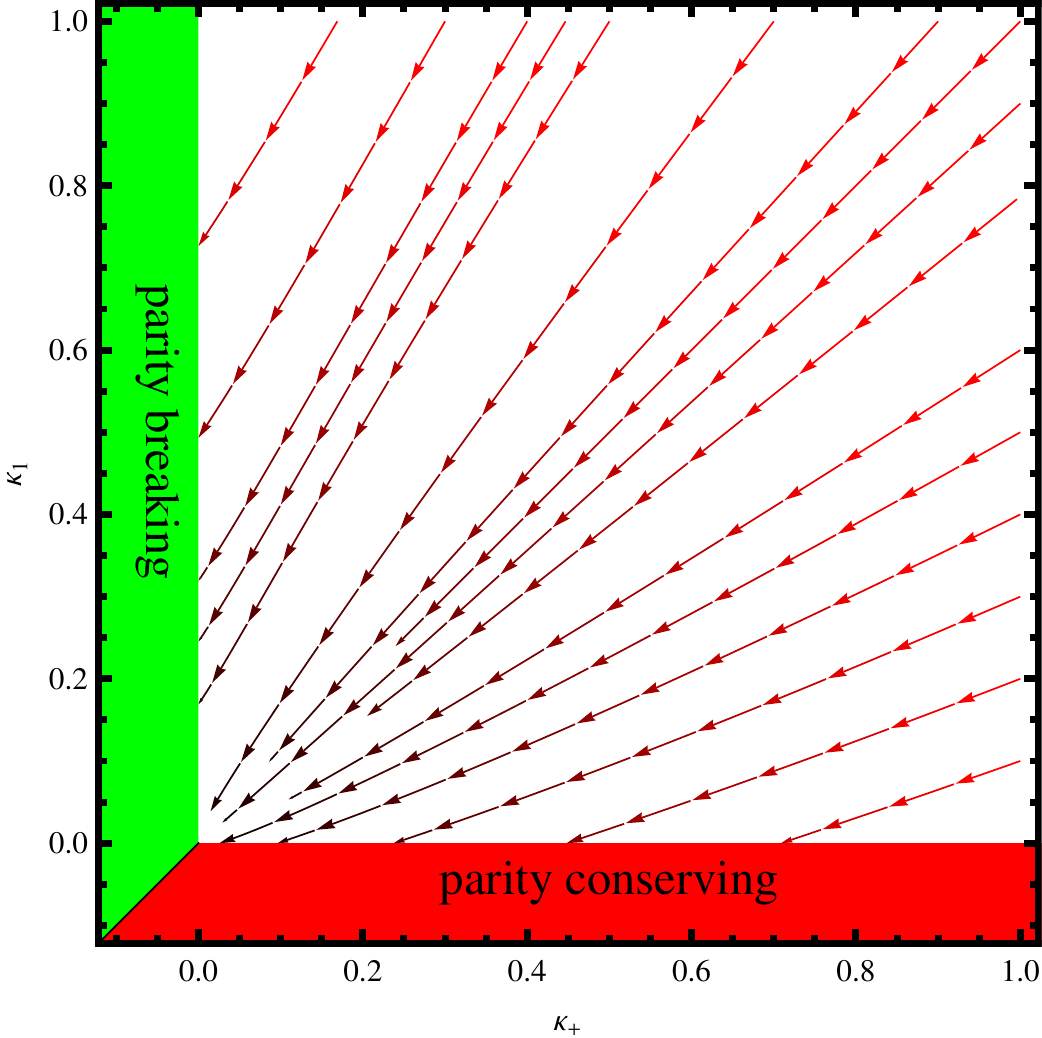}}
\hspace{1 cm}
\subfigure[Modification of RG flow by gauge boson
contributions.\label{fig:kappa2}]{\includegraphics[width=7.5 cm]{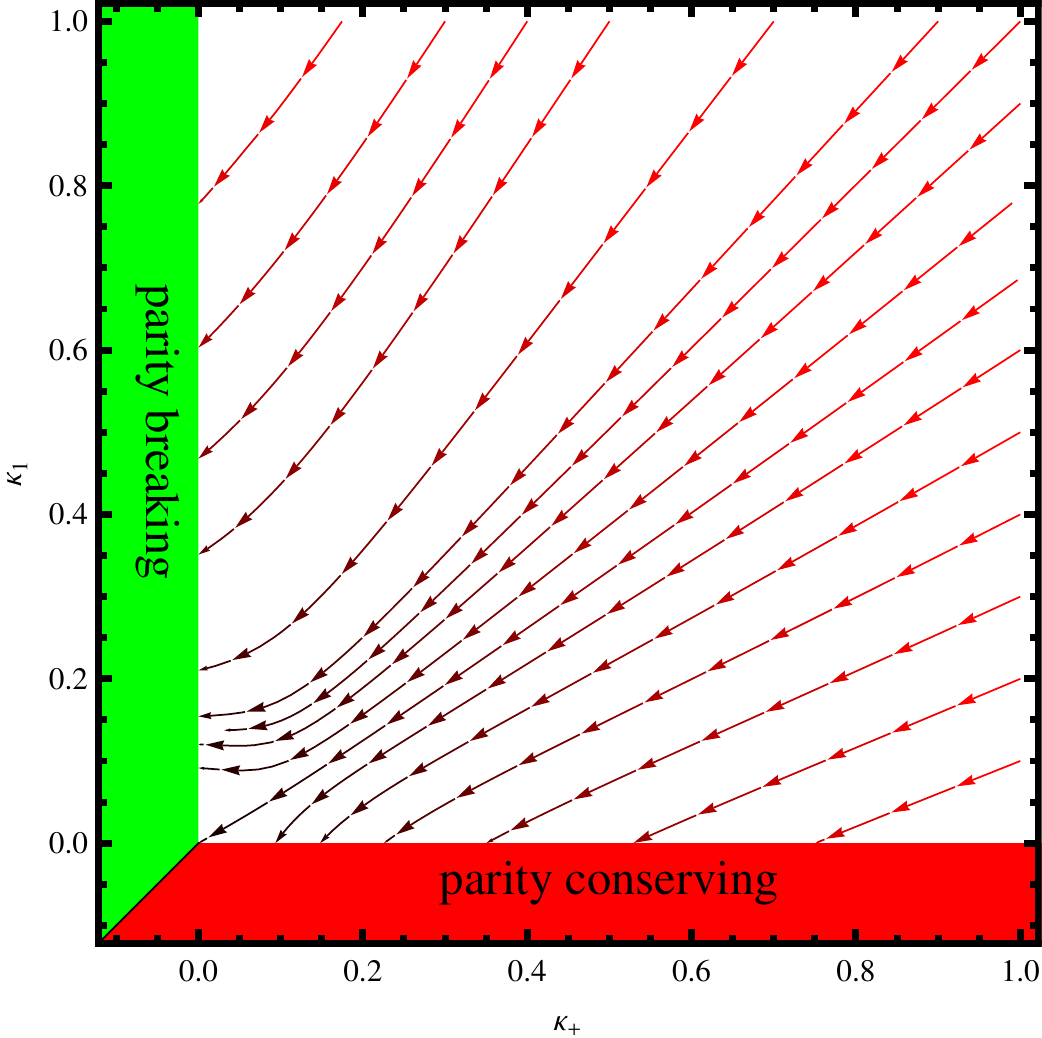}}
\hspace{1 cm}
\subfigure[Color Code]{\includegraphics[height=7.5cm]{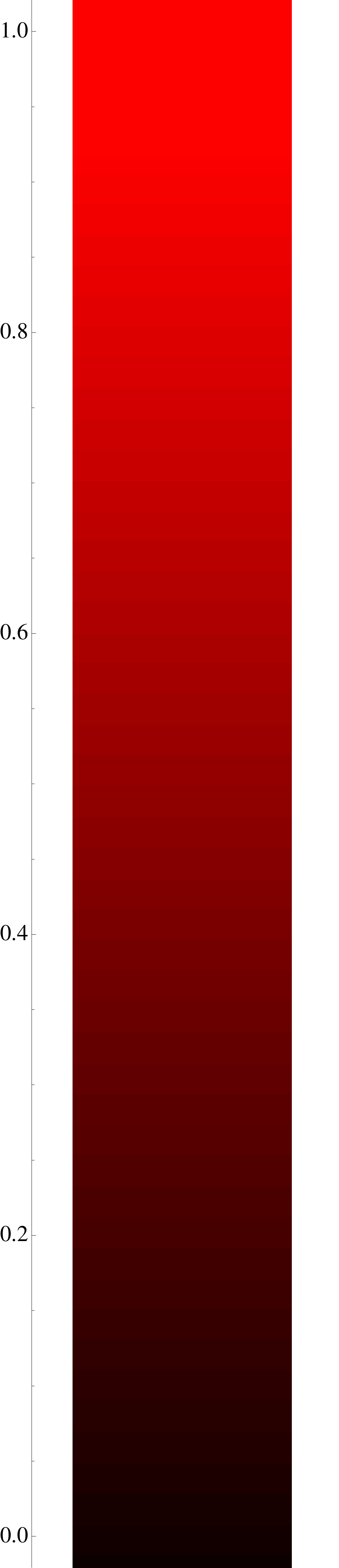}}
\hspace{1 cm}
\subfigure[RG flow in the small scalar coupling regime, where the gauge boson contributions dominate.\label{fig:kappa3}]{\includegraphics[width=7.5 cm]{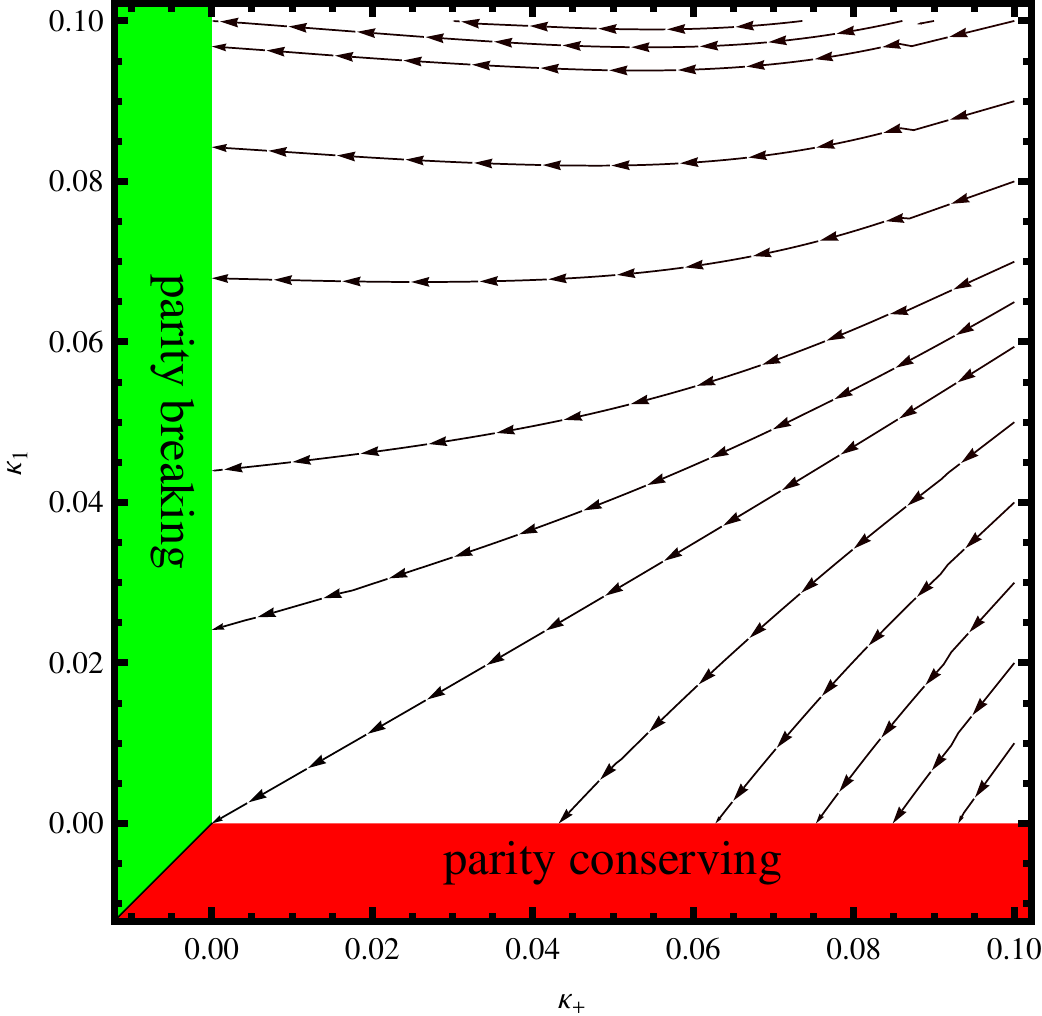}}
\caption{For small intermediate couplings between the doublet and bidoublet
  sectors, the main correction to RG flow in the  $\kappa_+$--$\kappa_1$ plane is coming from the gauge
  boson contributions. In \Figref{fig:kappa1}, the RG evolution towards lower energies
  without gauge bosons contributions is shown. As discussed in
  \Secref{sec:flatdirectionsofdoublet}, for any starting values with
  $\kappa_2>0$ the parity conserving GW condition $\kappa_1=0$ is reached
  eventually and for $\kappa_2<0$ the evolution tends towards the maximally
  parity-breaking solution $\kappa_+=0$. The effects of the gauge boson
  contributions are shown in \Figref{fig:kappa2}. For simplicity,
  the gauge couplings have been fixed to the values at $M_Z$, as they become
  relevant only at low energy scales. Following the stream lines,
  it can be seen that even for positive starting values of $\kappa_2$, the parity
  violating minimum might be reached. This region is depicted in  \Figref{fig:kappa3}. The gauge boson contributions deflect
  the RG evolution away from the vanishing coupling fixed-point. As the
  applicability of the GW formalism requires a sufficiently large $\kappa_1$ (see the discussion following \Eqref{eq:effectivecoupl}), this shows that the GW treatment is sufficient for a large portion of parameter space.}\label{fig:kappaflow}
\end{figure}

It is essential to study the RG flow to see whether a hypersurface described by
a GW condition is reached and which one is reached firstly. This decides if a symmetry breaks and how it breaks by this mechanism. Once the flat direction emerges at
tree level, the symmetry is broken and the subsequent running of the couplings is described
by the broken theory. The initial conditions of the RG equations and mainly the
beta functions of $\kappa_1$ and $\kappa_+$\footnote{All beta functions are summarized in \Appref{app:betafunctions}.}
\begin{subequations}
\begin{align}
8\pi^2\beta_{\kappa_1}&=5 \kappa_1^2+3 \kappa_+^2+
\left(\kappa_1-\frac34\left(g_1^2+\frac32 g_2^2\right)\right)^2-\frac{3}{16}
g_1^4-\frac32 g_1^2 g_2^2-\frac98 g_2^4+2 f_1^2+4 \beta_1^2\\
8 \pi^2 \beta_{\kappa_+}&=3 \kappa_1^2
+4\kappa_+^2+\left(\kappa_1-2\kappa_+\right)^2+\left(\kappa_+-\frac34 \left(
    g_1^2+\frac32 g_2^2\right)\right)^2 - \frac{3}{16} g_1^4 -\frac{21}{16}
g_1^2g_2^2 -\frac{63}{64} g_2^4 +4 f_1^2+4 \beta _1^2
\end{align}
\end{subequations}
determine the relevant GW hypersurface.
  Note, that $\beta_{\kappa_1}$ and $\beta_{\kappa_+}$ are
 positive as long as the gauge couplings are sufficiently small.
This means that coming from a higher scale, $\kappa_+$ will decrease until the
GW condition is satisfied and $\kappa_1$ will decrease unless the gauge couplings are
large.
As we are interested in a parity-breaking minimum, i.e. a solution for which the GW condition
$\kappa_+=0$ is fulfilled, we need $\kappa_2<0$ at the symmetry breaking
scale. This is realized if either $\kappa_2<0$ is also fulfilled at the Planck
scale or a small positive value of $\kappa_2$ at the Planck scale evolves to a
negative one:
\begin{equation}
 8\pi^2\beta_{\kappa_2}=3\left(\kappa_2-\frac14\left(g_1^2+\frac32g_2^2\right)\right)^2+3\kappa_2^2+8\kappa_1\kappa_2-\frac{3}{16}\left(g_1^4+2g_1^2g_2^2+\frac32
   g_2^4\right) +2f_1^2\; .
\end{equation}
In  \Figref{fig:kappaflow}, the RG flow towards lower energies in the
$\kappa_+$--$\kappa_1$ plane is depicted. The gauge boson contributions have the
effect of deflecting the couplings away from the point of vanishing couplings
and also of increasing the region of parameter space that leads to a maximally
symmetry breaking solution.

The logarithmic renormalization group running of the couplings naturally creates
 a large hierarchy between the LR breaking scale and the Planck scale.
In order to illustrate the hierarchy we show a numerical example. If $\kappa_1(M_{Pl})=1$ and $\kappa_2(M_{Pl})=-0.68$,
the GW condition is fulfilled at $\mu_{GW}\approx5.8 \TeV$ with
$\kappa_1(\mu_{GW})=-\kappa_2(\mu_{GW})=0.46$. From the minimum condition
\Eqref{eq:mincond} we can determine $\VEV{\phi}\approx10.4\TeV$, and find
$v_R=\frac{\VEV{\phi}}{\sqrt{2}}\approx 7.4 \TeV$. Symmetry breaking then
results in three heavy gauge bosons of $\SU{2}_R$, four real components of
$\oDt$ with mass $m^2=2 \kappa_1 v_R^2\approx \left(7.1 \TeV\right)^2$ and a
scalon with mass
\begin{equation}
m_s^2 = \frac{3 g_1^4+ 6 g_1^2g_2^2+9 g_2^4+64\kappa_2^2}{64\pi^2} v_R^2\approx \left(906\GeV\right)^2\;.
\end{equation}
Note that there is no fermionic contribution, as the fermions do not couple directly to the doublets.

\section{Combined LR and EW Symmetry Breaking}
\label{sec:RealisticModel}
In a next step we aim at the full LR and EW symmetry breaking sequence. At
first, we will discuss how the bidoublet can be included into the discussion and
subsequently we demonstrate how the GW conditions can be solved for the general Higgs
potential. Fermion masses and FCNCs are treated in
\Secref{sec:fermionmasses}. Finally, the RG evolution is outlined and we argue
that the GW method is applicable.

\subsection{Discussion of the Bidoublet Part}
In the last section, we have seen that by a proper choice of the renormalization point for $\kappa_2<0$ we can always change the renormalized couplings in such a way that the potential has a flat direction that maximally breaks parity. We will now extend this result to the case in which the bidoublet acquires a non-vanishing VEV and therefore breaks electroweak symmetry. For $\kappa_1=-\kappa_2$, the potential along the direction $\xi=N_2 \phi$ is given by
\begin{equation}
 \mathcal{V}_{\mathrm{GW}}=A \xi^4+B \xi^4 \ln \frac{\xi^2}{\mu_{GW}^2}
+\frac12\left( \beta_1  \tr \Bd^\dagger \Bd+f_1 \left(\Gamma\left[\Bd,\Bd^\dagger\right]  \right)_{33}   \right)\xi^2,
\end{equation}
with $A$ and $B$ being the $\ko{g^4}$ quantities defined in \Eqref{eq:AandB}.
To be able to decide if radiative symmetry breaking happens in this direction, the additional terms stemming from the bidoublet should not be larger than the loop contribution. Assuming all couplings to be of order $g^2$, one therefore needs
\begin{align}
 \frac{\tr \Bd^\dagger \Bd}{v_R^2}\lesssim\ko{g^2}.
\label{eq:2HDMconstitencycondition}
\end{align}
In this limit, it is possible to treat the left-right-symmetry
breaking\footnote{Indeed, it is also possible to use the conventional discussion
  by Coleman and Weinberg~\cite{Coleman:1973jx}.} without including the
bidoublet sector of the theory.
The EWSB induced by the bidoublet VEV $\langle \Bd \rangle$ is triggered by terms like $\beta_1 v_R^2 \tr \Bd^{\dagger} \Bd$. Due to the assumed VEV hierarchy, those terms dominate all quantum fluctuations, which are therefore neglected.
The potential of the bidoublet after LR-symmetry breaking is given by:
\begin{equation}
\mathcal{V}= \lambda_1 (\varphi_1^{\dagger}\varphi_1 + \varphi_2^{\dagger}\varphi_2)^2 + 4 \lambda_2 (\varphi_2^{\dagger}\varphi_1 + \varphi_1^{\dagger}\varphi_2)^2 +
 4 \lambda_3 (\varphi_2^{\dagger}\varphi_1 - \varphi_1^{\dagger}\varphi_2)^2
 +\frac{2
\beta_1 - f_1}{2} v^2_R
 \varphi_1^{\dagger}\varphi_1
+\frac{2 \beta_1 + f_1}{2} v^2_R \varphi_2^{\dagger}\varphi_2
\label{eq:2hdm}
\end{equation}
where the bidoublet has been decomposed into
\begin{equation}
 \oBd=\frac{1}{\sqrt{2}}(- \I \sigma_2 \varphi_1^*,\varphi_2)
\end{equation}
such that both
\begin{equation}
 \varphi_1=\left(\begin{array}{c}{\phi_2^-}^*\\ -{\phi_1^0}^* \end{array} \right) \andone \varphi_2=\left(\begin{array}{c} \phi_1^+\\ \phi_2^0 \end{array} \right)
\end{equation}
are weak doublets with hypercharge $y=1/2$. To discuss the stability and the minimization of this special Two Higgs Doublet Model\footnote{See for example~\cite{Gunion:1989we} and references therein.} (2HDM), we closely follow the approach of Maniatis et. al.~\cite{Maniatis:2006fs}. In their work, they eliminate all spurious gauge dependent degrees of freedom by discussing the potential in terms of gauge invariant field bilinears. They observe that the hermitian $2\times2$ matrix
\begin{equation}
M=\left(\begin{array}{ll}
\varphi_1^{\dagger}\varphi_1 & \varphi_2^{\dagger}\varphi_1 \\
\varphi_1^{\dagger}\varphi_2  & \varphi_2^{\dagger}\varphi_2\\
 \end{array}\right)
\label{eq:defofM}
\end{equation}
contains all relevant terms that show up in the potential.  Since the matrix is hermitian, it can be written as $M=\frac12 K_{\mu}\sigma^{\mu}$, using the Pauli matrices $\sigma^\mu=(\mathbb{1},\bm{\sigma})$. The Minkowski-type\footnote{Under a change of basis $\begin{pmatrix} \varphi_1\\ \varphi_2\end{pmatrix}\rightarrow U\begin{pmatrix} \varphi_1\\ \varphi_2\end{pmatrix}$ with $U\in U(2)$, $K_0$ transforms as a singlet, while $\textbf{K}=(K_1,K_2,K_3)$ transforms as a vector of $U(2)$. The four vectors with the upper indices are given by $K^\mu=\eta^{\mu \nu}K_\nu$ with the usual Minkowski metric $\eta$.} four vector $K_\mu$ parameterizes the gauge orbit of the potential. Positive definiteness ($ {v^*}^T M v={\mid v_1^* \varphi_1+v_2^* \varphi_2\mid}^2 \ge 0$) leads to the conditions:
\begin{equation}
 K_0=\tr M \ge 0 ,\quad\quad K_{\mu}K^{\mu}=K_0^2-\textbf{K}^2=4 \det M \ge 0.
\end{equation}
The allowed range therefore corresponds to the forward light-cone of Minkowski space. Rewriting the potential \Eqref{eq:2hdm} in terms of $K_\mu$
\begin{equation}
 \mathcal{V}=K_{\mu} \xi^{\mu}+K_\mu E^{\mu \nu} K_\nu
\end{equation}
we can discuss the properties of the potential in a gauge invariant way. For our potential, the parameters are given by
\begin{align}
 \left(\xi^\mu\right)&=2 \left(
\begin{array}{c}
 \frac{\beta_1 }{2} \\
 0 \\
 0  \\
 -\frac{f_1 }{4}
\end{array}
\right)v_R^2&\andone
\left(E^{\mu \nu}\right)&=\left(
\begin{array}{cccc}
 \lambda_1 & 0 & 0 & 0 \\
  0& 4 \lambda_2 & 0 & 0 \\
 0 & 0 & -4 \lambda_3 & 0 \\
 0 & 0 & 0 & 0
\end{array}
\right).
\label{eq:ourpotential}
\end{align}

\subsubsection{Stability}
To discuss the stability of the potential, we define the rescaled vector $\tilde{k}=K/K_0=(1,\textbf{k})$, for $K_0>0$.  $K^2\geq0$  implies $|\textbf{k}|\leq 1$.  From the form of the quartic part of the potential
\begin{equation}
\mathcal{V}_4 = K_0^2 \left( E^{00}+2 E^{0i} \tilde{k}_i+\tilde{k}_i E^{ij} \tilde{k}_j \right)=K_0^2 \left(  \lambda_1 + 4 \tilde k_1^2 \lambda_2 - 4 \tilde k_2^2 \lambda_3\right)=: K_0^2 J_4(\textbf{k})\;,
\end{equation}
it is then clear that for the potential to be stable (in the strong sense), the
condition \mbox{$J_4(\textbf{k})>0$} has to be fulfilled. Since the domain is compact, the global minimum of $J_4(\textbf{k})$  will be among the stationary points of the potential. The stationary points on the inside of the domain can be obtained by taking the gradient with respect to $\textbf{k}$, while those on the boundary are obtained by employing the Lagrange multiplier method. The condition that $J_4(\textbf{k})$ has to be larger than zero for each of those points, implies the following stability conditions on our potential:
\begin{align}
\lambda_1+4\lambda_2&>0,&
\lambda_1-4\lambda_3&>0,&
\lambda_1&>0\;.
\end{align}

\subsubsection{Minimization}
In order to find the global minimum of the potential, we have to find all stationary points within and on the lightcone. Stationary points within the lightcone have to fulfill
\begin{align}
 0=\frac{\partial}{\partial K_\mu}\mathcal{V}=2 E^{\mu \nu} K_\nu+\xi^\mu
\end{align}
with $K_\mu K^\mu>0$. Looking at \Eqref{eq:ourpotential}, we see that for $f_1\neq0$, this inhomogeneous linear equation does not have a solution for the potential we are interested in.\\
The stationary points on the lightcone are obtained by the method of Lagrange
multipliers and hence are given by the stationary points of the function
\begin{align}
F\left( K,u\right)&=\mathcal{V}-u K_{\mu}K^{\mu}.
\end{align}
Differentiation yields
\begin{equation}
\left(E^{\mu \nu}-u \eta^{\mu \nu} \right)K_\nu=-\frac{1}{2}\xi^\mu\quad\mathrm{as\ well\ as}\quad K_\mu K^\mu=0\quad\mathrm{and}\quad K_0>0
\end{equation}
For our set of parameters, no solution with  \mbox{$\det(E-u \eta)= 0$} exist. For regular values of $u$ with \mbox{$\det(E-u \eta)\ne 0$}, the solution is given by
\begin{align}
K_\mu(u)&=-\frac{1}{2}\left(E-u \eta \right)^{-1}_{\mu \nu}\xi^\nu
\end{align}
and the Lagrange multiplier is obtained from the constraint
\begin{align}
0&=K_\mu(u)K^\mu(u)=\frac{1}{4}\xi_\rho \left(E-u \eta \right)^{-1\,\rho \mu}\left(E-u \eta \right)^{-1}_{\mu \nu}\xi^\nu =:-{\tilde{f}'}(u)\;.
\end{align}
Only positive Lagrange multipliers correspond to minima, the global minimum is determined from the largest zero of $\tilde{f}'(u)$(see~\cite{Maniatis:2006fs}) . For our potential, the function $\tilde{f}'(u)$ is given by

\begin{equation}
\tilde f^\prime(u)=\frac{1}{16} v_R^4 \left(\frac{f_1^2}{u^2}-\frac{4 \beta _1^2}{\left(u-\lambda _1\right)^2}\right)
\end{equation}
which has the zeros $u_+=\frac{f_1 \lambda _1}{f_1-2 \beta _1}$ and
$u_-=\frac{f_1 \lambda _1}{f_1+2 \beta _1}$ and the field bilinears $K_\mu$ are given by
\begin{equation}
\left(K_\mu(u_\pm)\right)=\left(
\begin{array}{c}
 \frac{-\beta _1 }{2 \left(\lambda _1-u_\pm\right)} \\
0\\
 0\\
 \frac{f_1}{4 u_\pm}
\end{array}
\right)v_R^2
=\left(\begin{array}{c}1\\0\\0\\\pm1
  \end{array}
\right)\frac{\pm f_1-2 \beta
  _1}{4 \lambda _1}v_R^2\; .
\end{equation}  Before we go on to discuss which minimum is the global one, we
firstly have to consider another constraint which the global minimum has to fulfill.
\label{sec:minimization}

\subsubsection{EWSB}
While the correct breaking pattern $\SU2_L\times U(1)_Y\rightarrow U(1)_{em}$ is
obtained automatically in the standard one-doublet model, this is not generally
the case for the two doublet model. It is argued in~\cite{Maniatis:2006fs}, that
the $\U{1}_{em}$ conserving minimum lies on the boundary $K_\mu K^\mu=0$.
To be sure that there is a non-trivial minimum, the potential has to decrease as one goes away from the the origin:
\begin{align}
 0>\frac{\partial \mathcal{V}}{\partial K_0}\big|_{K_0=0}=\xi^\mu\tilde{k}_\mu=\xi_0-\bm{\xi}^T \textbf{k}>\xi_0-|\bm{\xi}| |\textbf{k}|\;,
\end{align}
Here, we have used that $K_0=\varphi_1^\dagger \varphi_1+\varphi_2^\dagger \varphi_2$ is a gauge and basis-independent measure of the distance from the origin of field space.
Since $K^2\ge0$  implies $\bm{k}^T\bm{k}\le1$, this gives the bound
\begin{align}
 \xi^0<|\bm{\xi}|,
\end{align}
which for our potential means
\begin{align}
 2 \beta_1<&\left|f_1\right|\;.
\label{eq:EWSB condition}
\end{align}
Since
$K_0=\varphi_1^\dagger\varphi_1+\varphi_2^\dagger\varphi_2$ and
$K_3=\varphi_1^\dagger\varphi_1-\varphi_2^\dagger\varphi_2$, it is clear that
for $u_+$ we have $\varphi_2=0$ and thus $\tan \beta=0$ and for $u_-$ we get
$\varphi_1=0$ and thus $\tan \beta=\infty$.
Using that the global minimum is given by the largest Lagrange multiplier and that $\varphi_i^\dagger\varphi_i>0$ shows that for $f_1>0$ the global minimum is given by $u_+$ and for $f_1<0$ the global minimum is given by $u_-$. \\
In \Eqref{eq:2HDMconstitencycondition}, we have seen that the way in which we have approached the problem in this section using the Two Higgs Doublet model requires a hierarchy between the electroweak scale and the LR breaking scale. The result of the minimization shows that
\begin{align}
\ko{g^2}\stackrel{!}{\gtrsim}\frac{\VEV{\tr \Bd^\dagger \Bd}}{v_R^2}=\frac{\VEV{\varphi_{1}^\dagger \varphi_{1}}+\VEV{\varphi_{2}^\dagger \varphi_{2}}}{v_R^2}=\frac{\kappa^2+{\kappa^\prime}^2}{ v_R^2}=\frac{|f_1|-2 \beta _1}{4 \lambda _1}
\end{align}
this condition generally requires some (fine-)tuning between the parameters of
the potential. Note that this is not the case for the the GW condition
$\kappa_+=0$, as this was related to the free choice of the renormalization
scale.

\subsection{Flat Directions of the General Higgs Potential}\label{sec:symmetriclimit}
\label{sec:flatdirections}
The  Higgs potential \Eqref{eq:VHiggssolo} can be minimized analytically. We use the GW condition
\begin{align}
 0&=\frac{\partial \mathcal{V}}{\partial \alpha}\Big|_{N_i=n_i}= -8 n_3^2 n_4^2 \sin\alpha \cos \alpha
\end{align}
to classify all different solutions. Clearly, these equations can only be fulfilled if either $n_3=0$, $n_4=0$ or $\alpha=0,\frac{\pi}{2}$.\\
The solutions Ia and Ib are given by $\alpha=0$ and $\alpha=\frac{\pi}{2}$, respectively. As the remaining equations are invariant under the transformation $(\alpha,\lambda_2,\lambda_3)\rightarrow(\frac{\pi}{2}-\alpha,-\lambda_3,-\lambda_2)$, the solutions of type Ib can be straightforwardly obtained from the solutions of type Ia. \\
The solutions IIa and IIb that are characterized by $n_3=0$ and $n_4=0$, respectively. Here we can also use the invariance of \Eqref{eq:GWconditionsfull} under the transformation $(n_3^2,n_4^2,f_1)\rightarrow(n_4^2,n_3^2,-f_1)$, and it is sufficient to discuss the case IIa.\\
Under the assumption that both $n_1$ and $n_2$ are non-vanishing, we can find another equation that can be used to classify the various solutions:
\begin{align}
0= \frac{1}{n_1}\frac{\partial \mathcal{V}}{\partial N_1}\Big|_{N_i=n_i}-\frac{1}{n_2}\frac{\partial \mathcal{V}}{\partial N_2}\Big|_{N_i=n_i}= \kappa_2\left(n_1^2-n_2^2\right)
\end{align}
Assuming $\kappa_2\neq0$,  all solutions therefore have to be either parity conserving in the doublet
sector -- meaning $n_1^2=n_2^2$ -- or maximally parity violating, meaning
$n_1\cdot n_2=0$. As the parity violating solutions are connected by the
replacement $n_1^2\rightarrow n_2^2$, it is sufficient to focus on the case
$n_1=0$. We will use an index $P$ and $\cancel{P}$ to distinguish between the
parity even and parity odd solutions in the doublet sector.\\
The remaining Eqs. \eqref{eq:GWconditionsfull} for the case $n_1^2=n_2^2$ can be mapped onto the equations for the case $n_1=0,n_2\neq0$ by the replacements $(n_1^2,n_2^2,\kappa_+)\rightarrow (n_2^2/2,n_2^2/2,\kappa_1)$. It is therefore sufficient to discuss the solutions with $n_1=0$ and $n_2\neq0$ as the other solutions can be obtained by those replacements.\\
All solutions (see \Tabref{table:flatdirections}) can therefore be obtained from the solutions $\mathrm{Ia}_{\cancel{P}}$ and $\mathrm{IIa}_{\cancel{P}}$ which will be discussed in the following.
\begin{table}[htb]
\begin{center}
\begin{tabular}[t]{l|c|c|c|c|c|c|c}
&GW condition& $\frac{n_1^2}{n_2^2}$&$n_1^2+n_2^2$&$\frac{n_3^2}{n_4^2}$&$n_3^2+n_4^2$&$\alpha$&$\frac{n_3^2+n_4^2}{n_1^2+n_2^2}$  \\\hline
$\mathrm{Ia}_{\cancel{P}}$&\multirow{2}{*}{$\left.\begin{array}{c}\kappa_+\\\kappa_1\end{array}\right\}=
 \frac{\beta_1^2}{2 (\lambda_1+4 \lambda_2)}-\frac{f_1^2}{32 \lambda_2}$} & $0$&\multirow{2}{*}{$\frac{2(
   \lambda_1+4 \lambda_2)}{2 \lambda_1+8  \lambda_2-\beta_1}$}&\multirow{2}{*}{$\frac{4\left(2\beta_1+f_1\right)\lambda_2+f_1\lambda_1}{4\left(2\beta_1-f_1\right)\lambda_2-f_1\lambda_1}$}&\multirow{2}{*}{$\frac{-\beta_1}{2\lambda_1+8\lambda_2-\beta_1}$}&$\multirow{2}{*}{0}$&\multirow{2}{*}{$
\frac{-\beta_1}{2 \lambda_1+8 \lambda_2} $}\\
$\mathrm{Ia}_{P}$&&$ 1$&&  &  &  &\\ \hline
 $\mathrm{Ib}_{\cancel{P}}$&\multirow{2}{*}{$\left.\begin{array}{c}\kappa_+\\\kappa_1\end{array}\right\}=
  \frac{\beta_1^2}{2 (\lambda_1-4 \lambda_3)}+\frac{f_1^2}{32 \lambda_3}$}&$0$& \multirow{2}{*}{$\frac{2 (\lambda_1-4 \lambda_3)}{2 \lambda_1-8 \lambda_3-\beta_1}$}&\multirow{2}{*}{$\frac{4\left(2\beta_1+f_1\right)\lambda_3-f_1\lambda_1}{4\left(2\beta_1-f_1\right)\lambda_3+f_1\lambda_1}$}&\multirow{2}{*}{$\frac{-\beta_1}{2\lambda_1-8\lambda_3-\beta_1}$}&\multirow{2}{*}{$ \frac{\pi}{2}$}&\multirow{2}{*}{$ \frac{-\beta_1}{2\lambda_1-8 \lambda_3}$}\\
 $\mathrm{Ib}_{P}$ &&$ 1$&&  &  &  &\\ \hline
Ic&$\lambda_1=-4\lambda_2 $&$\frac{0}{0}$&$ 0 $&$ 1$&$1 $&$ 0 $&$ \infty $\\\hline
Id&$\lambda_1=4\lambda_3 $&$\frac{0}{0}$&$ 0 $&$ 1 $&$1$&$ \frac \pi 2 $&$ \infty $\\\hline\hline
%%%
$\mathrm{IIa}_{\cancel{P}}$&\multirow{2}{*}{$\left.\begin{array}{c}\kappa_+\\\kappa_1\end{array}\right\}= \frac{(2 \beta_1-f_1)^2}{8 \lambda_1} $}&$0$&\multirow{2}{*}{$ \frac{4 \lambda_1}{-2 \beta_1+f_1+4 \lambda_1}$}&\multirow{2}{*}{$\infty$}&\multirow{2}{*}{$\frac{2 \beta_1-f_1}{2 \beta_1-f_1-4 \lambda_1}$}&\multirow{2}{*}{$-$}&\multirow{2}{*}{$ \frac{f_1-2 \beta_1}{4 \lambda_1} $}
\\
$\mathrm{IIa}_{{P}}$&&$1$&&&  & &\\\hline
$\mathrm{IIb}_{\cancel{P}}$&\multirow{2}{*}{$\left.\begin{array}{c}\kappa_+\\\kappa_1\end{array}\right\}= \frac{(2 \beta_1+f_1)^2}{8 \lambda_1}$}&$0$&\multirow{2}{*}{$ \frac{4 \lambda_1}{-2 \beta_1-f_1+4  \lambda_1}$}&\multirow{2}{*}{$ 0$}&\multirow{2}{*}{$\frac{2 \beta_1+f_1}{2 \beta_1+f_1-4 \lambda_1}$}&\multirow{2}{*}{$-$}&\multirow{2}{*}{$\frac{-f_1-2 \beta_1}{4 \lambda_1} $}
\\
$\mathrm{IIb}_{P}$&&$1$&&&&&\\\hline
 $\mathrm{IIc}_{\cancel{P}}$&\multirow{2}{*}{$\left.\begin{array}{c}\kappa_+\\\kappa_1\end{array}\right\}=0$}&$0$& \multirow{2}{*}{$1$}&\multirow{2}{*}{$\frac{0}{0}$}&\multirow{2}{*}{$0$}&\multirow{2}{*}{$-$}&\multirow{2}{*}{$0$}\\
 $\mathrm{IIc}_{P}$ &&$ 1$&&  &  &  &\\\hline
IId&$  \lambda_1=0 $&$\frac{0}{0}$&$ 0 $&$ 0 $&$1  $&$ - $&$ \infty $\\\hline
IIe&$  \lambda_1=0 $&$\frac{0}{0}$&$ 0 $&$ \infty $&$1  $&$ - $&$ \infty $\\
\end{tabular}
\caption{The flat directions of the model. Note, that for each maximally parity violating solution in the
  doublet sector, that is
  given by a condition on $\kappa_+$, there is another
  solution that is parity conserving given by the same constraint under the
  replacement $\kappa_+$ going to $\kappa_1$. Furthermore for every solution
  there is another one that is connected by parity. Parity interchanges
  $n_1$ and $n_2$, while leaving the other fields invariant. The angle $\theta$
  is left undetermined for all solutions, while the angle $\alpha$ is determined
  for the solutions of type I.\label{table:flatdirections}}
\end{center}
\end{table}

\mathversion{bold}
\subsubsection{Flat Direction $\mathrm{Ia}_{\cancel{P}}$}
\mathversion{normal}
For the case $n_1=\alpha=0$, the equations
\begin{subequations}
\begin{align}
0&=\frac{1}{n_3}\frac{\partial \mathcal{V}}{\partial
  N_3}\Big|_{N_i=n_i}-\frac{1}{n_4}\frac{\partial \mathcal{V}}{\partial
  N_4}\Big|_{N_i=n_i}=  - \left(\frac12 f_1 n_2^2+8 \lambda _2 \left(n_2^2+2 n_3^2-1\right)\right)\andone\\
0&=\frac{1}{n_2}\frac{\partial \mathcal{V}}{\partial
  N_2}\Big|_{N_i=n_i}=\frac14\left(f_1+2 \beta _1\right)
\left(-n_2^2-n_3^2+1\right)+\frac14 n_3^2 \left(2 \beta
   _1-f_1\right)+\frac12 n_2^2 \left(\kappa _1+\kappa _2\right)
\end{align}
\end{subequations}
can be easily solved to give
\begin{align}
n_2^2&=-\frac{32 \beta _1 \lambda _2}{f_1^2+32 \lambda _2 \left(-\beta _1+\kappa _1+\kappa
   _2\right)}, & n_3^2=\frac{f_1^2+2 f_1 \beta _1+32 \left(\kappa _1+\kappa _2\right) \lambda _2}{2 f_1^2+64
   \lambda _2 \left(-\beta _1+\kappa _1+\kappa _2\right)},&&n_4^2&=1-n_2^2-n_3^2
\end{align}
As before, the condition $\mathcal{V}\Big|_{N_i=n_i}=0$ gives an additional condition on the couplings:
\begin{equation}
\boxed{ \mathrm{Ia}_{\cancel{P}}$ \ :\  $\;$ $\kappa_+=\frac{\beta_1^2}{2(\lambda_1+4 \lambda_2)}-\frac{f_1^2}{32 \lambda_2}}
\end{equation}
In this direction, both neutral components of the bidoublet have non-vanishing vacuum expectation values. The ratio between the right-handed scale and the electroweak scale is given by
\begin{equation}
 \frac{\kappa^2+{\kappa^\prime}^2}{v_R^2}=-\frac{\beta_1}{2(\lambda_1+4\lambda_2)}
\end{equation}
and the ratio between the both bidoublet VEVs is given by
\begin{align}
 \frac{{\kappa}^2}{{\kappa^\prime}^2}&=\frac{8 \beta _1 \lambda _2+f_1 \left(\lambda _1+4 \lambda _2\right)}{8 \beta _1 \lambda _2-f_1 \left(\lambda
   _1+4 \lambda _2\right)}\approx -1 +32 \frac{\lambda_2}{f_1} \frac{\kappa^2+\kappa'^2}{v_R^2}+\ko{\left(\frac{\kappa^2+\kappa'^2}{v_R^2}\right)^2}\; .
\end{align}
Note, that ${\kappa}^2/{\kappa^\prime}^2<0$ to leading order which is in
  contradiction to the definition of $\kappa,\kappa^\prime>0$. Hence, the
next-to-leading order term has to compensate the leading order term in
$\frac{\kappa^2+\kappa'^2}{v_R^2}$ otherwise this is not a flat direction. This clearly requires some fine-tuning and is thus disfavored. Furthermore, this VEV ratio is related to the hierarchy between the top and the bottom mass, as we will show
in \Secref{sec:fermionmasses}.
If we want to explain this hierarchy
${\kappa}^2/{\kappa^\prime}^2\sim\left(m_t/m_b\right)^2\approx 40^2$ and assume
$\left(\kappa^2+{\kappa^\prime}^2\right)/v_R^2\approx\left(174 \GeV/5
  \TeV\right)^2$, we clearly need fine-tuning. This type of solution is therefore disfavored and we will thus focus on the solutions of type II.\\
Using the invariance under $(\alpha,\lambda_2,\lambda_3)\rightarrow(\frac{\pi}{2}-\alpha,-\lambda_3,-\lambda_2)$, we obtain solution Ib with the GW condition $\kappa_+=\beta_1^2/(2\lambda_3+8 \lambda_2)+f_1^2/(32 \lambda_3)$. For this solution, CP is spontaneously broken. In the limit $\lambda_1+4\lambda_2=0$ and $\lambda_1-4 \lambda_3=0$, these solutions smoothly turn into the solutions $\mathrm{Ic}_{\cancel{P}}$ and $\mathrm{Id}_{\cancel{P}}$ with the corresponding GW conditions $\lambda_1+4\lambda_2=0$ and $\lambda_1-4 \lambda_3=0$.

\mathversion{bold}
\subsubsection{Flat Direction $\mathrm{IIa}_{\cancel{P}}$}
\mathversion{normal}
The solution IIa for the case $n_1=n_4=0$ can be easily obtained from the condition
\begin{align}
0&=\frac{1}{n_3}\frac{\partial \mathcal{V}}{\partial N_3}\Big|_{N_i=n_i}=
-\frac14\left(f_1-2 \beta _1\right) n_2^2+ \lambda _1 n_3^2 =\frac14 \left(2 \beta_1-f_1-4 \lambda_1 \right)n_2^2 + \lambda_1
\end{align}
to be $n_2^2=4 \lambda_1/(4 \lambda_1-(2 \beta_1-f_1))$. The condition
$\mathcal{V}\Big|_{N_i=n_i}=0$ gives the additional condition on the
couplings:
\begin{equation}
\boxed{\mathrm{IIa_{\cancel{P}}}:\; \kappa_+=\frac{\left(f_1-2 \beta_1\right)^2}{8 \lambda _1}}
\end{equation}
We have to calculate the scalar mass spectrum, in order to see if this solution is a minimum and not just an extremum.
We denote the excitation in the flat direction, the so-called \emph{scalon}, by $s$ and its orthogonal complement by $h$\footnote{Here, we define for a complex field $\phi$: $\phi_r=\sqrt{2}\; \mathrm{Re}\, \phi$ and $\phi_i=\sqrt{2}\; \mathrm{Im} \, \phi$.}:
\begin{align}
 \left(\begin{array}{c}s\\h\end{array}\right)=\left(\begin{array}{cc}n_2&n_3\\-n_3& n_2  \end{array} \right)\left(\begin{array}{c} {\chi_{R}^0}_r \\ {\phi_1^0}_r\end{array} \right)=\left(\begin{array}{cc}\cos \vartheta&\sin \vartheta\\-\sin \vartheta& \cos \vartheta  \end{array} \right)\left(\begin{array}{c} {\chi_{R}^0}_r \\ {\phi_1^0}_r  \end{array} \right)=O(\vartheta)\left(\begin{array}{c} {\chi_{R}^0}_r \\ {\phi_1^0}_r  \end{array} \right).
\end{align}
The mixing angle is given by the relation between the EW and the LR breaking scale:
\begin{align}
 \tan^2 \vartheta=\frac{\kappa^2}{v_R^2}&= \frac{f_1-2 \beta _1}{4 \lambda _1}.
\end{align}
The relative magnitude of the right-handed breaking scale versus the electroweak
breaking scale, which we call little hierarchy in the following, is therefore set by the relative strength of the intermediate
couplings $\beta_1$ and $f_1$ to the quartic bidoublet coupling
$\lambda_1$. This is the case for all flat directions with non-vanishing VEVs
for both the bidoublet and the doublet. To obtain a reasonable hierarchy between
these scales therefore requires some fine-tuning, which is generic in
LR-symmetric potentials (see e.g.~\cite{Deshpande:1990ip}).\\
While the scalon mass vanishes at tree level, the state $h$ gets a mass of:
\begin{align}
 m_h^2=\frac12 ( f_1-2 \beta_1)\VEV{\phi}^2
\end{align}
with $\VEV{\phi}^2=({\VEV{{\chi_{R}^0}_r}}^2+{\VEV{{\phi_1^0}_r }}^2)=2(v_R^2+\kappa^2)$. Phenomenology requires the mixing angle to be small and we can therefore expand:
\begin{align}
 m_h^2\approx 4 \lambda_1 \; v_R^2 \tan^2 \; \vartheta  = 4 \lambda_1 \; \kappa^2.
\end{align}
Since the mixing angle is small, this field will play the role of the
Standard Model Higgs particle and we can apply the direct search bound
\begin{align}
 m_h^2\approx 4 \lambda_1 (174 \GeV)^2 \gtrsim (114 \GeV)^2 \Rightarrow \lambda_1 \gtrsim 0.11.
\end{align}
The other particle masses are given by
\begin{subequations}
\begin{align}
m_{\sigma_1}^2&=m_{\sigma_2}^2=\frac{f_1}{2} \VEV{\phi}^2&&\approx f_1 v_R^2 \\
m_{{\chi_L^0}_{r}}^2&=m_{{\chi_L^0}_{i}}^2=-\frac{4 \kappa _2 \lambda _1}{f_1-2 \beta _1+4 \lambda _1}\VEV{\phi}^2&&\approx-2\kappa_2 v_R^2 \\
m_{{\chi_L^-}_{r}}^2&=m_{{\chi_L^-}_{i}}^2= \frac{-f_1^2+2 f_1 \beta _1+8 \kappa _2 \lambda _1}{-2 f_1+4 \beta _1-8 \lambda _1}\VEV{\phi}^2=m_{{\chi_L^0}_{r}}^2+f_1\kappa^2&&\approx-2 \kappa_2 v_R^2\\
m_{{\phi_2^0}_r}^2&=  \frac{2 \left(-8 \beta _1 \lambda _2+f_1 \left(\lambda _1+4 \lambda _2\right)\right)}{f_1-2 \beta _1+4 \lambda _1}\VEV{\phi}^2&&\approx f_1 v_R^2+8 \lambda_2 \kappa^2\label{eq:FCNCmassRe}  \\
m_{{\phi_2^0}_i}^2&= \frac{2 \left(f_1 \left(\lambda _1-4 \lambda _3\right)+8 \beta _1 \lambda _3\right)}{f_1-2 \beta _1+4 \lambda _1}\VEV{\phi}^2 &&\approx f_1 v_R^2-8 \lambda_3 \kappa^2\label{eq:FCNCmassIm}
\end{align}
\end{subequations}
and the Goldstone bosons are given by
$
\pi_1, \pi_2, {\phi_2^-}_r,  {\phi_2^-}_i, {\phi_1^0}_i\,\,\mathrm{and}\,\, {\chi^0_R}_i,
$
with
\begin{align}
 \left(\begin{array}{c}\pi_1\\ \sigma_1\end{array}\right)&=O(-\vartheta)\left(\begin{array}{c} {\chi_{R}^-}_i \\ {\phi_1^+}_i  \end{array} \right)&&\mbox{and}& \left(\begin{array}{c}\pi_2\\\sigma_2\end{array}\right)&=O(\vartheta)\left(\begin{array}{c} {\chi_{R}^-}_r \\ {\phi_1^+}_r  \end{array} \right)\;.
\end{align}
The scalon mass is given by
\begin{equation}
m_s^2\approx \left(\frac{3 g_1^4+6 g_2^2 g_1^2+9 g_2^4+64 \left(\kappa
    _2^2+\beta_1^2\right) }{64 \pi ^2}+\Order{\frac{\kappa^2}{v_R^2}}\right)
v_R^2\; .
\end{equation}
If the smallness of the EW scale with respect to the LR breaking scale is due to
a cancellation of $f_1$ against $2 \beta_1$, meaning $(f_1-2 \beta_1)\ll f_1,
\beta_1$, all of these masses are of the order of the LR breaking scale and only
the Higgs particle $h$ has a mass of the EW scale. For this solution, only one of the neutral bidoublet components obtains a VEV. To be a be a minimum, all particle masses have to be positive. This
clearly implies $f_1>0$ and $\kappa_2<0$. There is another solution
($\mathrm{IIb}_{\cancel{P}}$) with the constraint $ \kappa_+=\left(f_1+2 \beta_1\right)^2/(8 \lambda _1)$ that leads to the same vacuum expectation
values, with $\kappa$ replaced by $\kappa^\prime$ and $f_1$ and $-f_1$
interchanged. These solutions precisely correspond to the solutions we have
found in \Secref{sec:minimization}.
The solution found in \Secref{sec:flatdirectionsofdoublet} is recovered from
solution IIa if $2\beta_1-f_1\rightarrow 0$.
In the limit $\lambda_1=0$ these solutions smoothly turn into the solutions $\mathrm{IId}_{\cancel{P}}$ and $\mathrm{IIe}_{\cancel{P}}$ with the GW condition $\lambda_1=0$.

Like in \Secref{sec:flatdirectionsofdoublet}, the GW conditions of the different flat directions can be translated
in stability conditions of the potential
\begin{equation}
\begin{split}
 &\min\left(\kappa_+,\,\kappa_++\frac{f_1^2}{32\lambda_2},\,\kappa_+-\frac{f_1^2}{32\lambda_3}\right)>0\,,\;\;
 \min\left(\kappa_1,\,\kappa_1+\frac{f_1^2}{32\lambda_2},\,\kappa_1-\frac{f_1^2}{32\lambda_3}\right)>0\,,\\
 &\lambda_1>0,\;\;
 \lambda_1+4\lambda_2>0 \andone
 \lambda_1-4\lambda_3>0\; .
\end{split}
\end{equation}

\subsection{Fermion Masses and FCNCs}\label{sec:fermionmasses}
As the Higgs boson in the Standard Model, in the minimal LR symmetric model the bidoublet plays the double role
of providing masses to the fermions and breaking electroweak symmetry. The Yukawa couplings
\begin{equation}
\mathcal{L}_\mathrm{Yuk}=\I {{Y_\Qu^+}}^{ij}\overline{\Qu}_i\frac{1+\Gamma}{2}
\Bd \Qu_j+\I{{Y_\Lp}^-}^{ij}\overline{\Lp}_i\frac{1-\Gamma}{2} \Bd \Lp_j
+\hc=
-{{Y_\Qu^+}}^{ij} \overline{\oQu}_i\oBd {\oQuD}_j
-{{Y_\Lp}^{-\dagger}}^{ij}\overline{\oLp}_i\tilde\oBd {\oLpD}_j
+\hc
\label{eq:Z4Yukawas}
\end{equation}
lead to the mass matrices
\begin{equation}\label{eq:wrongquarkmasses}
M_u=\frac{\kappa}{\sqrt{2}}{Y^+_\Qu}=\frac{\kappa}{\kappa^\prime}M_d,\hspace{2 cm}  M_l=\frac{\kappa}{\sqrt{2}}
{Y^-_\Lp}=\frac{\kappa}{\kappa^\prime}M_\nu \; .
\end{equation}
which are clearly not realistic as they would lead to vanishing mixing angles of
the quark and lepton mixing matrices and give phenomenologically wrong mass relations. By
introducing small $\mathbb{Z}_4$ breaking Yukawa couplings
\begin{equation}
\mathcal{L}_\mathrm{Yuk}^{\cancel{\mathbb{Z}_4}}=\I {{Y_\Qu^-}}^{ij}\overline{\Qu}_i\frac{1-\Gamma}{2}
\Bd \Qu_j+\I{{Y_\Lp}^+}^{ij}\overline{\Lp}_i\frac{1+\Gamma}{2} \Bd \Lp_j
+\hc=
-{{Y_\Qu^-}}^{ij} \overline{\oQu}_i\tilde\oBd {\oQuD}_j
-{{Y_\Lp}^{+\dagger}}^{ij}\overline{\oLp}_i\oBd {\oLpD}_j
+\hc
\end{equation}
one can in principle avoid these problems. Note that the largest $\mathbb{Z}_4$
breaking coupling would have to be of the order $\frac{m_b}{m_t}\sim
\ko{1\%}$. This will not induce large $\mathbb{Z}_4$ breaking scalar couplings,
as it can be shown using the beta functions of the full theory without the
$\mathbb{Z}_4$ symmetry which are given in \Appref{sec:Z4breaking}.\\
In the phenomenologically interesting case of condition $\mathrm{IIa}_{\cancel{P}}$ with $\kappa^\prime=0$, the mass matrices are given by
\begin{equation}
M_u=\frac{\kappa}{\sqrt{2}} {Y^+_\Qu},\hspace{2 cm} M_d=\frac{\kappa}{\sqrt{2}}
{Y^-_\Qu},\hspace{2cm}M_\nu=\frac{\kappa}{\sqrt{2}} {Y^+_\Lp},\hspace{2 cm} M_l=\frac{\kappa}{\sqrt{2}}
{Y^-_\Lp}\;
\end{equation}
which can be fitted to the experimental data, but there is no specific reason for
the smallness of neutrino masses or the flavor structure in general.

However, there is a general problem which has to be addressed in the context of minimal left-right symmetric models, namely the issue of FCNCs. While for the case $\kappa^\prime=0$ the neutral scalar $\phi_1^0$ is responsible for electroweak symmetry breaking and plays the role of the SM Higgs boson, the second scalar $\phi_2^0$ contained in the bidoublet has flavor changing interactions at tree level:
\begin{align}
 \mathcal{L}_{FCNC}=\overline{D}_L V^\dagger Y_u^{diag} V D_R \phi_2^0+\hc\;,
\end{align}
where $Y_u^{diag}=Y_\Qu^{+diag}$ and $V$ are the up-type Yukawa matrix in the mass basis and the CKM matrix, respectively. This leads to an effective $\Delta S=2$ Lagrangian~\cite{Mohapatra:1983ae,Ecker:1983uh,Gilman:1983bh,Gilman:1983ce,Pospelov:1996fq,Zhang:2007da}
\begin{align}
  \mathcal{L}_{\Delta S=2}=\frac{1}{M^2 \kappa^2}\left( \sum_{j=u,c,t}V^*_{jd}m_jV_{js}  \right)^2\left[(\overline{d}\gamma_5 s)^2-(\overline{d} s)^2 \right].
\end{align}
Here the mass of the real and imaginary parts of $\phi_2^0$ has been assumed to
be equal, $m({\phi_2^0}_r)=m({\phi_2^0}_i)=M$. As $M$ is the only free parameter
in this formula, one can obtain the bounds $M>15 \TeV$ from the $K_L$--$K_S$ mass
splitting and $M>25 \TeV$ from $B_s-\overline{B}_s$
mixing~\cite{Zhang:2007da}. This translates into a bound on the right-handed
scale in the multi-10 TeV range via the relation $M^2\approx f_1 v_R^2$ (see
\Eqqref{eq:FCNCmassRe}{eq:FCNCmassIm}) and thus precludes any discovery at the LHC and reintroduces a (small) hierarchy problem. One therefore either avoids this problem or lives with a certain amount of fine-tuning.\\
Note that this FCNC issue is independent of the VEV configuration of the bidoublet and
unavoidable in the minimal left-right symmetric models with a bidoublet. Models
with a low right-handed scale therefore either have to contain a mechanism for
giving a large mass to the FCNC Higgs particle while keeping the SM Higgs light, as
required by electroweak precision data, or they have to generate the light fermion
masses without the bidoublet.  Models that fall into the second category
are the so called Alternative Left-Right Symmetric
Models~\cite{Davidson:1987mh,Rajpoot:1987ji} that introduce vector-like
iso-singlet states which lead to see-saw masses for all fermions. Variants of these models can lead to a phenomenologically viable fermion mass spectrum and may avoid the FCNC problems.
\\
Motivated by the fact that the top quark mass is practically given by the
electroweak scale and much heavier than all other quarks, it seems plausible
that only the top mass is generated by the bidoublet and that a different mechanism generates the other masses. This idea can be
easily incorporated into our model. If we replace the $\mathbb{Z}_4$ symmetry by
a $\mathbb{Z}_{3_L}\times\mathbb{Z}_{3_R}$ symmetry in the spirit of~\cite{Babu:1990fr}
\begin{align}
\oBd&\stackrel{\mathbb{Z}_{3_R}}{\longrightarrow} \omega\, \oBd,& Q_{R,i}&\stackrel{\mathbb{Z}_{3_R}}{\longrightarrow} \omega^* Q_{R,i+1\,\rm{mod}\,3},& Q_{L,i}\stackrel{\mathbb{Z}_{3_L}}{\longrightarrow}  Q_{L,i+1\,\rm{mod}\,3}
\end{align}
with $\omega$ being the cubic root of unity, we find
$\lambda_2=\lambda_3=0$ and no additional scalar coupling with respect to the
$\mathbb{Z}_4$ case. There is furthermore only one allowed Yukawa coupling:
\begin{equation}
 -\mathcal{L}_{t}=\lambda_t \left(\overline{Q}_{L,1}+\overline{Q}_{L,2}+\overline{Q}_{L,3} \right)\oBd  \left( Q_{R,1}+Q_{R,2}+Q_{R,3}\right)
\end{equation}
which results in a mass matrix of rank one and thus only the top quark mass has a
mass at tree level. The remaining fermion masses might then be generated
radiatively along the lines of~\cite{Balakrishna:1988bn,Ma:1989tz,Dobrescu:2008sz}. Since the bidoublet only
couples to the top quark, it does not generate any FCNCs.  In the RG analysis in the next section we assume that the top mass is due to the Yukawa interaction with the bidoublet and that the FCNC problem is solved.
In addition to generating a realistic fermionic spectrum, any extension of the minimal model presented here should also break the accidental symmetry
\mbox{$\AccSymm: \Dt\rightarrow e^{\I \beta \Gamma}\Dt$}, because there is a
remnant symmetry \mbox{$\AccSymm: \oDt\rightarrow\mathrm{e}^{\I\beta}\oDt$} in
the preferred vacuum with $\VEV{\chi_L}=0$. This remnant symmetry leads to a stable particle
$\oDt^0$ which is in conflict with direct detection
experiments~\cite{Ahmed:2008eu} because of its vector-like coupling to the Z
boson. Hence, the accidental symmetry has to be broken either completely or to a
$\mathbb{Z}_2$ parity in order to generate a mass splitting between the real and imaginary parts of $\oDt^0$. This splitting then kinematically forbids the coupling to the Z boson(see e.g.~\cite{TuckerSmith:2001hy}).
Note that in the Alternative Left-Right Symmetric Models, $\AccSymm$ is completely broken by the Yukawa coupling to the additional vector-like fermions.

\subsection{Renormalization Group Evolution}
\label{sec:RGfull}
In \Secref{sec:RGparity}, we have demonstrated that the parity breaking solution
is obtained for a large part of parameter space. This is still valid in the full
theory, since the mixing parameters $\beta_1$ and $f_1$ are supposed to be small
which can be seen from the $\beta$-functions shown in
\Appref{app:betafunctions}.
The scalar coupling $f_1$ is multiplicatively renormalized, because it is the only coupling which breaks the separate \Spin{4}
transformations of $\Dt$ and $\Bd$ to the diagonal subgroup.
As it has been pointed out in \Secref{sec:flatdirections}, the hierarchy between
EWSB and LR symmetry breaking scale is determined by the smallness of the parameters mixing $\Dt$ and $\Bd$, namely $\beta_1$ and $f_1$
and the quartic couplings of $\Bd$. At tree level, the vanishing of $\beta_1$ and $f_1$ can be
explained by a symmetry. However, the coupling $\beta_1$ is generated via gauge boson loops and therefore the
little hierarchy cannot be explained by a (nearby) symmetry in this model.

In order to illustrate the scalar mass spectrum, we explicitly describe one
scenario where left-right symmetry can be spontaneously broken by radiative
corrections under the assumption, that the contribution of the additional physics to explain the
down-type quark as well as neutrino masses is negligible. The main fermionic
contribution stems from the top quark. The Higgs mass is
fixed to $m_h=120\GeV$ and the VEV of the RH doublet to
$v_R=3\TeV$ which leads to gauge boson masses consistent with the experimental
bounds~\cite{Amsler:2008zzb}\footnote{We do not take into account the bound on
  the right-handed scale recently obtained from the neutron electric dipole moment~\cite{Xu:2009nt} in the minimal left-right symmetric model with triplets, because we did not consider down-type quark masses and it does not
  directly apply to Alternative Left-Right Symmetric Models where the right-handed down quark is not the $\SU{2}_R$ partner of the up quark.}. Furthermore, we choose $\lambda_2=-\lambda_3=0.001$, $\kappa_1=0.2$ and $\beta_1=0.01$ which leads to
$\lambda_1\approx0.119$, $\kappa_2\approx-0.200$ and
$f_1=2.16\cdot 10^{-2}$ at the LR breaking scale. Those parameters evolve  to
\begin{align}
\lambda_1&\approx0.0285,&\lambda_2&\approx0.00637,&\lambda_3&\approx-0.0271,&\kappa_1&\approx0.325,&\kappa_2&\approx-0.102,\label{eq:boundaryPlanck}\\\nonumber
\beta_1&\approx0.0242,&f_1&\approx0.0119
\end{align}
at the Planck scale.
At the energy scale  $\mu_{GW}\approx 15  \TeV$ all parity-breaking conditions
are fulfilled. However, the potential has one flat direction of type IIa only,
since the other parity-breaking solutions are either unstable or correspond to
imaginary field values. The scalar masses are given by
\begin{align}
m_{{\chi_L^0}_{r}}^2=m_{{\chi_L^0}_{i}}^2\approx m_{{\chi_L^-}_{r}}^2=m_{{\chi_L^-}_{i}}^2&\approx  1900\GeV\nonumber\\
m_{\sigma_1}^2=m_{\sigma_2}^2&\approx 442\GeV\\\nonumber
m_{{\phi_2^0}_r}^2\approx m_{{\phi_2^0}_i}^2&\approx 441\GeV\; .
\end{align}
Hence, the spectrum is split in a light sector which mainly originates from the
bidoublet $\Bd$ and an heavy sector originating from the doublet $\Dt$.
The scalon mass
\begin{align}
 m_s= 294 \GeV
\end{align}
is suppressed compared to the other masses originating from the doublet $\Dt$ by
a loop factor. The mixing angle between the SM Higgs and the neutral CP-even component of $\chi_R$ is given by
\begin{equation}
\tan \vartheta=\frac{\kappa}{v_R}\approx0.06
\end{equation}
which should lead to an interesting phenomenology of this scenario.
The right-handed gauge boson masses are above 1 TeV:
\begin{align}
 M_{Z_R}&= 1640\GeV  &\mbox{and}&&  M_{W_R^{\pm}}&=1360 \GeV\; ,
\end{align}
but are still within the LHC reach. If it is possible to measure the the mixing
angle $\theta$ by the different branching ratios of the SM Higgs and the agent
of LR breaking $\chi_R^0$ one can test our model by comparing it to the ratio of
gauge boson masses:
\begin{align}
 \tan \vartheta=\frac{\kappa}{v_R}\approx \frac{m(W^\pm_L)}{m(W^\pm_R)}.
\label{eq:striking}
\end{align}
This would be a striking indication for conformal left-right breaking.

\subsection{Applicability}
\label{sec:applicability}
The small values of the couplings at the LR breaking scale call into question
the applicability of the GW method in this scenario, because all
couplings are assumed to be  of the order of the gauge coupling squared in~\cite{Gildener:1976ih} to ensure that the symmetry can only be broken in the flat
direction. However, this condition can be replaced by the weaker condition that
the potential  along an arbitrary direction $\phi_i=N_i \phi$
\begin{equation}
 V=\frac{\lambda_{\{N_i\}}}{4!} \phi^4
\label{eq:effectivecoupl}
\end{equation}
should have an effective coupling $\lambda_{\{N_i\}}\sim \ko{g^2}$. As the terms $\frac{\kappa_1}{2}(\overline{\Dt}\Dt)^2$ and $\lambda_1 (\tr \Bd^\dagger \Bd)^2$ can be written in real component fields as the squares of $\mathrm{O}(4)$-symmetric field bilinears
\begin{equation*}
\frac{\kappa_1}{2}(\overline{\Dt}\Dt)^2=\frac{\kappa_1}{4}\left(N_1^2+N_2^2+N_5^2+N_6^2 \right)^2\phi^4
\end{equation*}
and
\begin{equation*}
\lambda_1 (\tr \Bd^\dagger \Bd)^2= \frac{\lambda_1}2 \left(N_3^2+N_4^2+N_7^2+N_8^2 \right)^2\phi^4,
\end{equation*}
those terms give the same contribution in any direction in field space, and the
Gildener-Weinberg method can be applied as long as $12 \lambda_1\sim\ko{g^2}$
and $8 \kappa_1\sim\ko{g^2} $.
According to~\cite{Gildener:1976ih}, $g$ is defined as the gauge or Yukawa coupling
resulting in the largest contribution to the effective potential which is due to
the top Yukawa coupling $y_t \sim \ko{1}$ in our model. Hence, the quartic
scalar couplings $\lambda_1$ and $\kappa_1$ should be of order $\ko{0.1}$, which
is fulfilled. Therefore, the GW discussion is applicable.

\section{Summary and Conclusions}
\label{sec:summary}
In this work, we studied the symmetry breaking mechanism in the minimal conformally invariant left-right symmetric model. We were motivated to study such a scale-invariant model because of the recent work of Meissner and Nicolai, who have argued that classical conformal symmetry might be considered as a possible
alternative solution to the hierarchy problem within the framework of 4D quantum
field theory.  \\
Using the framework of Gildener and Weinberg~\cite{Gildener:1976ih}, we
calculated all flat directions and classified them according to symmetries of
the scalar potential. It turns out that there are only two inequivalent classes
(I and II) of such directions which are not connected by a symmetry or contained
as limit in another case. The solutions of type I only exist for a rather small
part of the parameter space that gives $v_R\gg\kappa$. Thus we focused on the
solutions of type II which result in $\kappa^\prime=0$.
We have discussed the phenomenologically most interesting flat direction
$\mathrm{IIa}_{\cancel{P}}$ in detail. The one-loop effective potential along
the flat direction stabilizes the scalon and determines its mass. All other scalar masses are either close to the LR symmetry
 or close to the electroweak symmetry breaking scale depending on the quartic
 scalar couplings. This splitting is due to the assumed smallness of $\beta_1$ and $f_1$ which describe the mixing
 between the bidoublet $\Bd$ and the doublet(s) $\Dt$. In the limit of vanishing gauge interactions, the smallness is natural
 because in the limit of vanishing $\beta_1$ and $f_1$ the conformal
 transformations of $\Bd$ and $\Dt$ become independent and the symmetry is
 enhanced~\cite{'tHooft:1980xb}. However, on the quantum level, the coupling $\beta_1$ is generated by
 gauge boson loops and the little hierarchy between electroweak and LR symmetry
 breaking scale cannot be explained.
The study of flat directions in the Gildener-Weinberg framework is complemented by a study
of the renormalization group equations which are summarized in \Appref{app:betafunctions} in order to determine the relevant flat direction for the
ground state of the model. Parity  is broken in a large fraction of parameter space and
 there is a region in parameter space which leads to the flat direction
$\mathrm{IIa}_{\cancel{P}}$, as it has been shown in \Seccref{sec:RGparity}{sec:RGfull}.
 Hence, a phenomenologically viable scalar mass spectrum can be obtained as well
 as up-type quark and charged lepton masses. However, the
 $\mathbb{Z}_4$ symmetric model is not a realistic model of fermion
 masses. Supplementing this model by small $\mathbb{Z}_4$ breaking Yukawa
 couplings one can obtain a realistic spectrum. The problem of this approach to
 fermion masses is, however, that whenever the up-type fermion
 masses are generated through the bidoublet, FCNCs constrain the right-handed scale to be above
 $20 \TeV$. This reintroduces a hierarchy problem and is a general feature of LR
 symmetric models. Since this work is focused on the scalar sector and the
 symmetry breaking sequence, we did not present a full theory of fermion masses
 but only alluded to various solutions that exist in the
 literature~\cite{Davidson:1987mh,Rajpoot:1987ji,Balakrishna:1988bn,Ma:1989tz,Dobrescu:2008sz}. However,
 one  might raise the question why the bidoublet is introduced in the first place. There are two answers to
 this question: Firstly, we strongly suspect that the large top mass is
 generated at tree level by the bidoublet Yukawa couplings, and secondly without the bidoublet one finds
 $v_L=0$ and therefore no EWSB. In Alternative Left-Right Symmetric Models~\cite{Davidson:1987mh,Rajpoot:1987ji}, the
 latter problem is usually addressed by introducing LR breaking scalar mass terms,
 which we do not consider as it would violate both conformal and LR symmetry.
Interestingly, one can introduce a discrete symmetry
$\mathbb{Z}_{3_L}\times\mathbb{Z}_{3_R}$, which results in the Yukawa coupling \mbox{$(\sum_i \overline Q_{L,i})\oBd(\sum_j
  Q_{R,j})$}, as it has been discussed in
  \Secref{sec:fermionmasses}. Hence, the up-type quark mass matrix has rank one
and therefore only the top mass is generated at tree level.\\
It should also be stressed that the scalar spectrum of the model shows
  some peculiar features that enable to distinguish the model from LR
  symmetric models that do not originate from a classically conformal
  theory. The two lightest scalars in the theory are the SM Higgs and the
  excitation in the flat direction of the potential (the scalon). The mixing angle
  $\vartheta$ between the two states should be experimentally accessible through
  the branching ratios of the various decays and can be compared to the mass-ratio of
  the left- and right-handed W bosons. The conformal breaking scenario presented
  in this paper predicts these to be equal, \mbox{$\tan \vartheta=\kappa/v_R\approx
  m(W^\pm_L)/m(W^\pm_R)$} (see \Eqref{eq:striking}). We believe that a
  confirmation of this prediction would constitute a striking indication for
  conformal left-right symmetry breaking. One should further note that this
  prediction is only possible due to the enlarged gauge group and marks an
  advantage of this model with respect to singlet extensions of the SM.

Let us comment further on the view of the hierarchy problem we take in
  this paper following the approach of Meissner and Nicolai. We assume that there is a UV finite
  theory of gravity that becomes classically conformally invariant in the flat
  space limit, which is taken to be the fundamental reason for the appearance of
  small mass scales in comparison to the Planck scale in nature. Since the
  Planck scale effects are communicated to the particle physics action only
  through the logarithmic running of the coupling constants, the generation of
  the hierarchy seems well motivated. It should be noted, however, that the
  stabilization of this hierarchy explicitly requires this special property of
  the quantum gravitational embedding and therefore requires "'something extra'
  beyond quantum field theory"~\cite{Meissner:2009gs}. Any embedding into a
  conventional quantum field theory at any scale between the LR breaking scale
  and the Planck scale would lead to quadratical corrections of the Higgs mass
  term of the order of the embedding scale. This is to be contrasted with
  supersymmetry, which does not require some special property of Planck scale
  physics to stabilize the weak scale, but where the generation of the hierarchy
  between the scales remains unexplained.

Concluding, classical conformal symmetry may have a role to play in generating the hierarchy between the Planck and electroweak scales. In this paper we studied a model exhibiting many generic features of conformal left-right breaking. It will be interesting to explore variants which systematically avoid FCNCs. It also seems very promising to study versions where the Higgs boson
is the pseudo Nambu-Goldstone boson of an additional symmetry, which would be a natural explanation for
the separation of the electroweak and left-right breaking scales.

\section*{Acknowledgements}
We would like to thank R.~N.~Mohapatra for useful discussions. This work was partly supported by the Sonderforschungsbereich TR 27
of the Deutsche Forschungsgemeinschaft.
\appendix

\section{Description of the Model}
\label{app:spin4}
As explained before, we use the isomorphism between $\SU{2}\times\SU{2}$ and \Spin{4} to express the minimal left-right symmetric model based on the gauge group $\SU{2}_L\times\SU{2}_R\times U(1)_{B-L}$ in terms of the various representations of \SO{4} of which \Spin{4} is the double covering group.  The quarks and leptons transform as spinors of \SO{4}:
\begin{align}
\Lp \rightarrow \exp\left(\frac{1}{2}\alpha_{AB}\Sigma^{AB}\right)\Lp=:S(A)\Lp
\end{align}
 where $A=1+\alpha$ is a general \SO{4} transformation. The explicit form of the generators  $\Sigma^{AB}$ can be obtained from the Clifford algebra $\left\{\Gamma^A,\,\Gamma^B\right\} = 2\delta^{AB}$ in the same way as for the Lorentz group. We use a hermitian representation given by
\begin{equation}
\Gamma^A=
\left(\begin{array}{cc}
0 & \sigma^A\\
\bar\sigma^A & 0 \\
\end{array}
\right).
\end{equation}
with $\sigma^A=(\bm{\sigma},\I \mathbb{1})$, $\overline{\sigma}^A=(\bm{\sigma},-\I \mathbb{1})$ and the Pauli matrices $\bm{\sigma}=(\sigma^1,\sigma^2,\sigma^3)$.
The generators of the \Spin{4} group are formed by $\Sigma^{AB}=\frac{1}{4}\left[\Gamma^A,\,\Gamma^B\right].$ We can define a chirality operator
\begin{equation}
\label{eq:chiralityOperator}
\Gamma=\Gamma^1\Gamma^2\Gamma^3\Gamma^4=\left(\begin{array}{cc}
1 & \\
& -1 \\
\end{array}
\right)
\end{equation}
which allows us to define projection operators on left-, and right-chiral
states
$
\mathbb{P}_{L/R}=\frac{1\pm\Gamma}{2}.
$
The standard left-handed $\SU{2}_L$ doublet $L_L=\left(\nu_L \  e_L \right)^T$ can be obtained from $\Lp$ by $L_L=P_L \mathbb{P}_L \Lp$ and $ L_R=P_R \mathbb{P}_R \Lp$ with the standard Lorentz group projection operators $P_L$ and $P_R$. The spinors further have to fulfill the condition
\begin{align}
 P_L \mathbb{P}_R \Lp =P_R \mathbb{P}_L \Lp=0\; .
\label{eq:spinorconstraint}
\end{align}
The chirality operator anticommutes with all $\Gamma$ matrices $\left\{\Gamma,\Gamma^A\right\}=0$
and therefore commutes with $\Sigma^{AB}$ and the charge conjugation
operator
\begin{equation}
\mathcal{C}=\Gamma^2\Gamma^4\; .
\end{equation}
The charge conjugation matrix $\mathcal{C}$ further satisfies
${\mathcal{C}\Gamma^A}^{T} \mathcal{C}^{-1}=\Gamma^A$
which ensures that
$\Sigma^{AB}=\mathcal{C}\Sigma^{AB*}\mathcal{C}^{-1}$.
Apart from the spinor representation, we use the vector representation which corresponds to the bidoublet in the minimal LR symmetric model. We use the Clifford
algebra to rewrite the vector $\phi_A$, that transforms as $\phi_A\rightarrow
A_{AB}\phi_B$, as the bidoublet matrix $\Bd=\phi_A \Gamma^A$ that transforms
according to
\begin{align}
 \Bd\rightarrow S(A)\Bd S^{-1}(A).
\end{align}
We further define a parity transformation $\mathbb{P}$ in complete analogy to
the Lorentz group. Under parity, a $\SO{4}$ spinor transforms as
\begin{subequations}
\begin{equation}
\mathbb{P}:\Dt\rightarrow \Gamma^4 \Dt
\end{equation}
and the bidoublet transforms as
\begin{equation}
\mathbb{P}:  \Bd\rightarrow -\Gamma^4 \Bd^\dagger \Gamma^4\; .
\end{equation}
\end{subequations}
The phases of the transformations have been chosen such that in terms of the
standard fields, we recover the transformation properties $\chi_L\leftrightarrow
\chi_R$ and $\oBd\leftrightarrow {\oBd}^\dagger$. The scalar fields $\oBd$ and $\Dt$
in components read
\begin{subequations}
\begin{align}
\Bd&=\left(\begin{array}{cc} 0 &
      \oBd \\ -\oBdD & 0\end{array}\right)
&
\oBd&=\frac{1}{\sqrt{2}}\left(\begin{array}{cc}
\phi_1^0 & \phi_1^+\\
\phi_2^- & \phi_2^0\\
  \end{array}\right)
\\
\Dt&=\left(\begin{array}{c}\oDt\\-\I\,\oDtD
  \end{array}\right)
&
\chi_{L/R}&=\left(\begin{array}{c}
\chi_{L/R}^0\\
\chi_{L/R}^-\\
\end{array}\right)
\end{align}
\end{subequations}
where the real components of the complex scalar fields
$\chi_{L/R}^0$, $\phi_i^0$, $\chi_{L/R}^-$, $\phi_1^+$ and $\phi_2^-$ are denoted by the indices $r$ and $i$ in
the form
\begin{equation}
\phi=\frac{\phi_r+\I\,\phi_i}{\sqrt{2}}\; .
\end{equation}

\section{Lagrangian}
\label{app:Lagrangian}
The gauge kinetic part is defined by
\begin{equation}
\mathcal{L}_\mathrm{gauge}=-\frac{1}{4}B_{\mu\nu}B^{\mu\nu}+\frac{1}{4}\tr
\mathbb{W}_{\mu\nu}\mathbb{W}^{\mu\nu}-\frac{1}{2}\tr G_{\mu\nu}G^{\mu\nu}\; .
\end{equation}
The field strength corresponding to the $\Spin{4}$ gauge connections is defined by
\begin{equation}
 [D_\mu,D_\nu]\big|_\mathbb{W}= \frac{g_2}{\sqrt{2}} \left(\partial_\mu \mathbb{W}_\nu-\partial_\nu \mathbb{W}_\mu \right)+\frac{g_2^2}{2}[\mathbb{W}_\mu,\mathbb{W}_\nu]=: \frac{g_2}{\sqrt{2}} \mathbb{W}_{\mu\nu}
\label{eq:Commgauge}
\end{equation}
and the $\SU{3}_C$ color gauge group as well as the $B-L$ symmetry are
described as usual. The $\Spin{4}$ field strength is defined by
\begin{equation}
\begin{split}
\mathbb{W}_{\mu\nu}&=\partial_{\mu}\mathbb{W}_\nu
-\partial_{\nu}\mathbb{W}_\mu+\frac{g_2}{\sqrt{2}}[ \mathbb{W}_{\mu},
\mathbb{W}_{\nu}]\\
&=\left[\partial_{\mu}W^{AB}_\nu
  -\partial_{\nu}W^{AB}_\mu+\frac{g_2}{\sqrt{2}}\left(W_{\mu}^{BE}W_{\nu}^{AE}-W_{\nu}^{BE}W_{\mu}^{AE}\right)\right]\Sigma^{AB}
\end{split}
\end{equation}
The fermionic kinetic term can be written
\begin{equation}
\mathcal{L}_{\rm{kin,fermion}}= \overline{\Lp}\I \cancel{D}^\Lp \Lp+\overline{\Qu}\I \cancel{D}^\Qu \Qu
\end{equation}
where the covariant derivatives are defined by
\begin{align}
D_\mu^\Lp&= \partial_\mu-\I \frac{1}{2} g_1 B_\mu+\frac{1}{2}\frac{g_2}{\sqrt{2}}W_{\mu}^{AB}\Sigma^{AB}\\
D_\mu^\Qu&=\partial_\mu+\I \frac{1}{6} g_1
B_\mu+\frac{1}{2}\frac{g_2}{\sqrt{2}} W_{\mu}^{AB}\Sigma^{AB}
+\I\frac{g_3}{2}G_{\mu}^m \lambda^m\; ,
\end{align}
$\lambda^m$ being the Gell-Mann matrices.
The scalar kinetic term is given by
\begin{equation}
\mathcal{L}_{\rm{kin,scalar}}= \overline{D_\mu\Dt} D^\mu\Dt+ \tr \left(D_\mu\Bd\right)^\dagger D^\mu \Bd
\end{equation}
Finally, the Higgs potential is shown in \Eqref{eq:VHiggssolo} and the Yukawa couplings in \Eqref{eq:Z4Yukawas}.

\section{Renormalization Group Equations}
\label{app:betafunctions}
Here, we collect all beta functions. The RG equations of the gauge couplings are described by
\begin{equation}
16\pi^2 \beta_{g_A}:=16\pi^2\mu\frac{\D g_A}{\D\mu}= b_A g_A^3
\end{equation}
with the coefficients
$(b_{\SU{3}_C},\,b_{\Spin{4}},\,b_{\U{1}_{B-L}})=(-7,\,-\frac{17}{6},\,3)$. The
$\beta$ functions of the remaining parameters in the Lagrangian are
\begin{subequations}
\begin{align}
\beta_{\beta_1}&=\frac{1}{256 \pi ^2}\left[-4 \beta_1 \left(-8 \beta_1+6 g_1^2+27 g_2^2-2 (20 \kappa_1+4 \kappa_2+40 \lambda_1+32 \lambda_2-32 \lambda_3+T_2)\right)+24 f_1^2+9 g_2^4\right]\\
\beta_{f_1}&=\frac{f_1}{64 \pi ^2} \left[16 \beta_1-6 g_1^2-27
  g_2^2+8\kappa_1+8\kappa_2+16 (\lambda_1-4 \lambda_2)+64 \lambda_3+2 T_2\right]\\
\beta_{\kappa_1}&=\frac{1}{512 \pi ^2}\left[\kappa_1\left(-96 g_1^2 -144 g_2^2 +576 \kappa_1+384  \kappa_2\right)+192 \kappa_2^2+256 \beta_1^2+128 f_1^2+24 g_1^4+12 g_1^2 g_2^2+9 g_2^4\right]\\
\beta_{\kappa_2}&=\frac{1}{512 \pi ^2}\left[\kappa_2\left(-96 g_1^2 -144 g_2^2 +512 \kappa_1 +384 \kappa_2\right)+128 f_1^2+12 g_1^2 g_2^2+9 g_2^4\right]\\
\beta_{\lambda_1}&=\frac{1}{128 \pi ^2}\left[\lambda_1\left(-72
g_2^2 +256\left(\lambda_1+ \lambda_2- \lambda_3\right)+8 T_2)\right) +1024\left( \lambda_2^2+\lambda_3^2\right)+32 \beta_1^2+8 f_1^2+9 g_2^4-4 T_4\right]\\
\beta_{\lambda_2}&=\frac{1}{512 \pi ^2}\left[\lambda_2\left(-288
g_2^2 +768 \lambda_1 +3072 \lambda_2+1024
\lambda_3+32  T_2\right)-8 f_1^2+3 g_2^4+2 T_{4}\right]\\
\beta_{\lambda_3}&=\frac{1}{256 \pi ^2}\left[\lambda_3(-144 g_2^2  +384 \lambda_1-512\lambda_2-1536 \lambda_3+16 T_2)+4 f_1^2-3
  g_2^4-  T_4\right]\\
\beta_{Y_{\Lp}^-} &= \frac{1}{64\pi^2}\left[(-6  g_1^2-9 g_2^2) Y_{\Lp}^-
+  Y_{\Lp}^- T_2
+4 {Y_{\Lp}^-}^3
\right]\\
\beta_{Y_{\Qu}^+}&=\frac{1}{64\pi^2} \left[(-\frac29  g_1^2-9 g_2^2-32g_3^2) Y_{\Qu}^+
+  Y_{\Qu}^+ T_2
+4 {Y_{\Qu}^+}^3
\right]
\end{align}
\end{subequations}
where we have used
\begin{subequations}
 \begin{align}
T_2& =\tr({Y_{\Lp}^-}^2+3{Y_{\Qu}^+}^2)\\
T_4&=\tr( {Y_\Lp^-}^4+3  {Y_\Qu^+}^4)\; .
 \end{align}
\end{subequations}

\section{Gildener Weinberg Conditions}
\begin{subequations}
\begin{align}
0=\frac{\partial \mathcal{V}}{\partial N_1}\Big|_{N_i=n_i}&= n_1 \left(-n_3^2 \left(f_1-2 \beta _1\right)+n_4^2 \left(f_1+2 \beta _1\right)+2 n_1^2 \left(\kappa _1+\kappa _2\right)+2 n_2^2 \left(\kappa _1-\kappa
   _2\right)\right) \\
 0=\frac{\partial \mathcal{V}}{\partial n_2}\Big|_{N_i=n_i}&=n_2 \left(-n_3^2 \left(f_1-2 \beta _1\right)+n_4^2 \left(f_1+2 \beta _1\right)+2 n_1^2 \left(\kappa _1-\kappa _2\right)+2 n_2^2 \left(\kappa_1+\kappa _2\right)\right) \\
0=\frac{\partial \mathcal{V}}{\partial N_3}\Big|_{N_i=n_i}&=
n_3 \left(4 \left(n_3^2+n_4^2\right) \lambda _1-\left(n_1^2+n_2^2\right)
  \left(f_1-2 \beta _1\right)\right)
+32  n_3 n_4 \left(n_3 \lambda _2 \cos ^2(\alpha )- n_4\lambda _3 \sin ^2(\alpha )\right) \\
 0=\frac{\partial \mathcal{V}}{\partial N_4}\Big|_{N_i=n_i}&=
n_4 \left(4
  \left(n_3^2+n_4^2\right) \lambda _1+\left(n_1^2+n_2^2\right) \left(f_1+2 \beta
    _1\right)\right)+32  n_3 n_4 \left(n_4 \lambda _2 \cos^2(\alpha )- n_3 \lambda _3 \sin ^2(\alpha )\right) \\
  0=\frac{\partial \mathcal{V}}{\partial \alpha}\Big|_{N_i=n_i}&= -8 n_3^2 n_4^2   \left(\lambda _2+\lambda _3\right) \sin (\alpha )\cos (\alpha ) \\
 0=\frac{\partial \mathcal{V}}{\partial \theta}\Big|_{N_i=n_i}&=0 \\
0=\mathcal{V}\Big|_{N_i=n_i}&=
n_4^2 \left(\left(n_1^2+n_2^2\right) \left(f_1+2 \beta _1\right)+4 n_3^2 \left(\lambda _1+4 \lambda _2-4
   \lambda _3\right)\right)+16 n_3^2 n_4^2 \left(\lambda _2+\lambda _3\right) \cos (2 \alpha )\nonumber\\&
+\left(n_1^2+n_2^2\right)^2 \kappa_1+\left(n_1^2-n_2^2\right)^2 \kappa _2+2 \left(n_3^4+n_4^4\right) \lambda _1-\left(n_1^2+n_2^2\right)
   n_3^2 \left(f_1-2 \beta _1\right) \\
 1&=n_1^2+n_2^2+n_3^2+n_4^2
\end{align}
\label{eq:GWconditionsfull}
\end{subequations}

\mathversion{bold}
\section{$\mathbb{Z}_4$ Breaking Terms}\label{sec:Z4breaking}
\mathversion{normal}

The $\mathbb{Z}_4$ breaking terms in the Higgs potential are
\begin{equation}
\mathcal{V}_{\cancel{\mathrm{Z}_4}}=\lambda_4\left(\tr \Bd
  \Bd^{\dagger}\right)\left(\tr\Bd\Bd+\tr\Bd^{\dagger}\Bd^{\dagger}\right)
+\beta _2\left(\tr\Bd\Bd+\tr\Bd^{\dagger}\Bd^{\dagger}\right)\overline{\Dt}\Dt
+\I
\beta_3\left(\tr\Bd\Bd-\tr\Bd^{\dagger}\Bd^{\dagger}\right)\overline{\Dt}\Gamma\Dt\; .
\end{equation}
In principle, there is also a kinetic mixing term $\tr D_\mu\Bd D^\mu \Bd$, which we choose to vanish by an appropriate basis
transformation. It is, however, generated radiatively. We take this into account
in the calculation of the $\beta$-functions by a rediagonalization of the
kinetic term analogously to the Yukawa couplings\footnote{In our calculation
  the off-diagonal terms do not arise at the one-loop level.}.
They lead to changes in the RG equations ($\beta_\lambda={\beta_\lambda}_{\mathbb{Z}_4}+\delta \beta_\lambda$) which are summarized as follows
\begin{subequations}
\begin{align}
\delta\beta_{\beta_1}=&\frac{1}{64 \pi ^2}\Big[32 (\beta_2^2- \beta_3^2)+4 \beta_2 (48 \lambda_4-T_-)-\beta_1 T_+\Big]\\
\beta_{\beta_2}=&\frac{1}{64 \pi ^2}\Big[\beta_2 \left(-6 g_1^2-27 g_2^2+2 (20
    \kappa_1+4 \kappa_2+8 \lambda_1+160 \lambda_2+32 \lambda_3+8 \beta_1
    + T_2 + T_+)\right)\nonumber\\
  &+ 48 \lambda_4\beta_1- T_-\beta_1\Big]\\
\beta_{\beta_3}=&\frac{\beta_3}{64 \pi ^2} \left(16 \beta_1-6 g_1^2-27
  g_2^2+8\kappa_1+40 \kappa_2+16 (\lambda_1-4 \lambda_2)-320 \lambda_3+2T_2+ 2 T_+\right)\\
\delta\beta_{f_1}=&\frac{f_1}{32 \pi ^2}  T_+\\
\delta\beta_{\kappa_1}=&\frac{2}{\pi ^2} \beta_2^2\\
\delta\beta_{\kappa_2}=&-\frac{2}{ \pi ^2} \beta_3^2\\
\delta\beta_{\lambda_1}=&\frac{1}{256 \pi ^2}\Big[-\mathcal{T}_+-\mathcal{T}_{3}+16 \left(48 \lambda_4^2-\lambda_4 T_-+\lambda_1 T_+\right)\Big]\\
\delta\beta_{\lambda_2}=&\frac{1}{512 \pi ^2}\Big[128 \beta_2^2+384 \lambda_4^2+32 \lambda_2 T_+-8 \lambda_4
T_- - \mathcal{T}_++\mathcal{T}_{3}\Big]\\
\delta\beta_{\lambda_3}=&\frac{1}{256 \pi ^2}\left[32 \beta_3^2+16\lambda_3 T_+- \mathcal{T}_4\right]\\
\beta_{\lambda_4}=&\frac{1}{64 \pi ^2}\left[\lambda_4\left(-36 g_2^2
    +192\lambda_1+768 \lambda_2 +4T_2 + 4 T_+\right) - 2\,T_- (\lambda_1 + 4 \lambda_2)+32\beta_1\beta_2+\frac12\mathcal{T}_-\right]\\
\beta_{Y_{\Lp}^+}=&\frac{1}{64\pi^2} \left[(-6  g_1^2-9 g_2^2) Y_{\Lp}^+
+Y_{\Lp}^+ (T_2+T_+)
+ Y_{\Lp}^-T_-
+4 {Y_{\Lp}^+}^3-2 \{{Y_{\Lp}^-}^2,Y_{\Lp}^+\}
\right]\\
\delta\beta_{Y_{\Lp}^-} =& \frac{1}{64\pi^2}\left[
Y_{\Lp}^- T_+
+ Y_{\Lp}^+ T_-
+4 {Y_{\Lp}^-}^3 -2  \{Y_{\Lp}^-,{Y_{\Lp}^+}^2\}
\right]\\
\delta\beta_{Y_{\Qu}^+}=&\frac{1}{64\pi^2} \left[
Y_{\Qu}^+ T_+
+ Y_{\Qu}^-T_-
+4 {Y_{\Qu}^+}^3-2 \{{Y_{\Qu}^-}^2,Y_{\Qu}^+\}
\right]\\
\beta_{Y_{\Qu}^-} =& \frac{1}{64\pi^2}\left[(-\frac29  g_1^2-9 g_2^2-32g_3^2) Y_{\Qu}^-
+Y_{\Qu}^- (T_2+T_+)
+ Y_{\Qu}^+ T_-
+4 {Y_{\Qu}^-}^3 -2  \{Y_{\Qu}^-,{Y_{\Qu}^+}^2\}
\right]
\end{align}
\end{subequations}
where we have used
\begin{subequations}
 \begin{align}
T_+& = \tr({Y_{\Lp}^+}^2+3 {Y_{\Qu}^-}^2)\\
T_-&=2\,\tr(Y_{\Lp}^-Y_{\Lp}^+)+6\,\tr(Y_{\Qu}^+Y_{\Qu}^-)\\
\mathcal{T}_{+}&=\tr {Y_\Lp^+}^4+4 \tr {Y_\Lp^+}^2{Y_\Lp^-}^2+2\tr \left({Y_\Lp^+}{Y_\Lp^-}\right)^2 +3\,(\Lp^\pm\rightarrow \Qu^\mp)\\
\mathcal{T}_{-}&=2\,\left(\tr {Y_\Lp^-}{Y_\Lp^+}^3+\tr {Y_\Lp^+}{Y_\Lp^-}^3 \right)+3\,(\Lp^\pm\rightarrow \Qu^\mp)\\
\mathcal{T}_3&=3 \tr {Y_\Lp^+}^4-4 \tr {Y_\Lp^+}^2{Y_\Lp^-}^2-2\tr \left({Y_\Lp^+}{Y_\Lp^-}\right)^2+3\,(\Lp^\pm\rightarrow \Qu^\mp)\\
\mathcal{T}_4&=\tr {Y_\Lp^+}^4-4 \tr {Y_\Lp^+}^2{Y_\Lp^-}^2+2\tr
\left({Y_\Lp^+}{Y_\Lp^-}\right)^2+3\,(\Lp^\pm\rightarrow \Qu^\mp)\; .
 \end{align}
\end{subequations}

\bibliographystyle{ArXiv-collref}
\bibliography{EWSB}

\end{document}